\documentclass{article}

\usepackage[applemac]{ inputenc}

\usepackage{mathrsfs}
\usepackage{amsmath}
\usepackage{amssymb}
\usepackage{multicol}
\usepackage{float}
\usepackage{makeidx}
\usepackage{layout}
\usepackage{array}
\usepackage{boxedminipage}
\usepackage{hyperref}
\usepackage{latexsym}
\usepackage{indent first}

\usepackage{subcaption,graphicx}

\usepackage{color}

\usepackage[sectionbib,square]{natbib}

\usepackage[normalem]{ulem}

\newcommand{\F}{\mathbf{F}}
 \newcommand{\C} {\mathcal{C}}

\title{Spontaneous critical layer formation and  robustness beneath rotational waves}
\author{M. V. Flamarion$^{1}$, A. Nachbin$^{2}$   and R. Ribeiro-Junior$^{3}$}

\date{}

\begin{document}

\maketitle

{\footnotesize
	\begin{center}
	$^{1}$ UFRPE/Rural Federal University of Pernambuco, UACSA/Unidade Acad{\^e}mica do Cabo de Santo Agostinho, BR 101 Sul, 5225, 54503-900, Ponte dos Carvalhos, Cabo de Santo Agostinho, Pernambuco, Brazil.
	
	$^2$ IMPA/National Institute of Pure and Applied Mathematics, Est. D. Castorina, 110, Rio de Janeiro, RJ, 22460-320, Brazil. 
	
	$^3$ UFPR/Federal University of Paran\'a,  Departamento de Matem\'atica, Centro Polit\'ecnico, Jardim das Am\'ericas, Caixa Postal 19081, Curitiba, PR, 81531-980, Brazil. 
	
	\end{center}
	
	}

\begin{abstract}
	Non-stationary rotational surface waves are considered, where the underlying current has constant vorticity. A study is presented on the robustness of a critical layer in the presence of a bottom topography, as well as on its spontaneous  formation for waves generated from rest. The restriction, from previous studies, to a traveling-wave formulation is removed  leading to a non-stationary set of equations.  In this setting streamlines are not necessarily  pathlines.
	Particle-trajectories are found evolving the respective submarine dynamical system with a  cloud 
	of tracers.  Pathlines are then visualized and  the respective submarine structures identified.  Robustness is illustrated through
	surface waves interacting with topographic undulations.  The respective Kelvin cat eye structure
	dynamically adjusts itself over the bottom topography without loosing its integrity.
	On the spontaneous formation of a Kelvin cat eye structure,  the surface is initially undisturbed and waves are generated  from either  
	the current-topography interaction or by a surface pressure distribution suddenly imposed. 
	Under the pressure forcing, an isolated Kelvin cat eye spontaneously forms. The two extreme critical points of the cat eye structure are connected with a stagnation segment, thus exhibiting a new form of a critical layer.
	

\end{abstract}

\section{Introduction and background}

Water waves propagating in the presence of a background flow is a problem of great current interest  both regarding applications as well as the mathematical  questions
that arise.
This topic is broad and therefore difficult to make a comprehensive overview of contributions.
For the interested reader we mention just a few recent references from which the bibliography can be useful. 
Regarding geophysical applications, \cite{Soontiens} studied  trapped internal waves over isolated topography,
in the presence of a background shear. The book by \cite{Buhler} presents several techniques relevant to GFD (Geophysical Fluid Dynamics) where an
example includes {``\it the interplay between large-scale Rossby waves and two-dimensional turbulence"}.  On the other extreme of the scientific spectrum,
we have many contributions in Analysis with rigorous theoretical results on the existence of surface waves in the presence of vorticity (namely rotational waves),
on the existence of stagnation points in the wave's moving frame, the respective 
Kelvin cat eye recirculation region below nonlinear (periodic, traveling) Stokes waves, 
among others. Many recent references  may be found through the work of Constantin and collaborators (\cite{ConstantinBook, ConstantinTrans,ConstantinActa}). 
Numerical studies with traveling waves and the respective  stationary submarine structures, such as stagnation points, start with the work of \cite{teles&peregrine}. More 
detailed numerical studies on the flow structure beneath traveling waves appeared recently, as for example that of  \cite{vasan&oliveras} and also \cite{JFM17}.
A numerical stability study for finite-amplitude steady rotational surface waves is presented by \cite{Kharif}.


Regarding wave-current interaction, special attention has been given to sheared currents of constant vorticity, a regime which 
permits the existence of stagnation points in the wave's moving frame.
These stagnation points are part of a recirculation region, displaying a Kelvin cat eye formation. Some theoretical and computational studies, such as those
mentioned above,  have considered 
different issues of interest related to particle trajectories beneath a rotational traveling wave.  Having traveling waves in mind, most studies are formulated 
in the wave's moving frame thus addressing stationary differential equations. The stationary submarine structure for  particle dynamics, for example, can be 
visualized through level-curves of the streamfunction.  

In the present study, in order to study the  robustness and the spontaneous  formation of the Kelvin cat eye structure and its associated critical layer, 
we need to consider a non-stationary set of equations. Even if the initial disturbance has data related to a traveling wave, the evolution equations are not
``aware" of this class of data and should (hopefully) evolve the initial wave profile as a pure translation. Unfortunately in the nonlinear case 
there is strong evidence that
traveling-wave data evolves as a stable wave of translation only for small amplitudes. \cite{Kharif} studied the stability of rotational waves and report an interesting
result that 
{\it ``increasing the shear reduces the growth rate of the most unstable sideband instabilities but enhances the growth rate of these quartet instabilities, 
	which eventually dominate the Benjamin-Feir modulational instabilities."}
To the best of our knowledge there are no articles with time dependent models and non-stationary waves 
studying the submarine Kelvin cat eye structure and its associated critical layer. The critical layer 
defines the boundary where the submarine flow (beneath the wave) is divided into a rightgoing and a leftgoing stream.   

As we remove the restriction of traveling waves,  our propagating  waves might change their  profiles as time evolves  
implying that the streamlines are no longer  pathlines. 
In this case, not only we have to compute the free surface conditions in time, but also we have to solve the particle-trajectories' dynamical system for a  cloud 
of tracers in order to visualize  the pathlines and the respective submarine structures.  

The {\it robustness} of the submarine structure is tested first by allowing an initial surface disturbance to propagate over the bottom topography. 
Due to the problem of 
nonlinear wave instability we perform our numerical study with linear waves, which has not yet been explored.
We consider  a modulated initial surface disturbance, where the Kelvin cat
eye structure is already present at time $t=0$. The initial wavetrain is chosen so that dispersive effects are very small and we observe effectively (namely to
a good approximation) a wave of translation. 
We will show that as 
the wave interacts with the topographic undulations, commensurable with its wavelength,  the Kelvin cat eye
dynamically adjusts and ``slides" over the topography without loosing its integrity. When the finite-length topography ends, the recirculation region is
back to its original cat eye formation. 
In the supplemental material a video shows the entire process.

A rapidly varying topography is also considered. In this case we cannot identify the Kelvin cat eye over the topography. 
The material curves, defined through groups of tracers, end up crossing each other.
Nevertheless, it is remarkable to observe the robust reconfiguration of the submarine structure, identified by
the respective group of tracers, as soon as the bottom becomes flat again.  

The novel results on the {\it  spontaneous} formation of the Kelvin cat eye structure are  considered through surface waves generated from rest.
The free surface is initially undisturbed when
surface waves start being generated by the current-topography interaction, or  by a localized (steady) surface pressure distribution 
suddenly applied at time $t=0^+$.  This case is motivated by the work of \cite{johnson}.
Under the pressure forcing two localized surface pulses are generated  and propagate  in 
opposite directions.  An isolated  Kelvin cat eye spontaneously forms under the downstream-propagating pulse. A
center (critical point) appears in the middle of the recirculation region.   The two extreme 
critical points of the cat eye structure are connected with a stagnation segment, thus exhibiting a new form of a critical layer. 
All these dynamical critical-layer scenarios described above have not been contemplated in the literature.  

The paper is organized as follows. In section 2 we present the mathematical formulation of the free surface Euler equations in the canonical domain,
a uniform strip where computations are more easily performed. The canonical
domain is defined through a conformal mapping. In  section 3 the numerical method is presented. We introduce the dynamical system for  particle trajectories 
in canonical coordinates. By not using a traveling-wave formulation this dynamical system is no longer autonomous and its vector field must be 
constantly updated. This update depends on  solutions of the Euler equations and is done through the potential component of the velocity field. Properties of
harmonic functions are used in order to write all  ``Euler-information" needed in terms of one-dimensional 
Fourier expressions. These are essentially Fourier-type operators acting on the Dirichlet (boundary) data. This framework leads to the 
numerical method described in section 3. The results are presented in section 4 and the conclusions in section 5. 

\section{Mathematical formulation} \label{formulacao}

We have a  two-dimensional incompressible flow of an inviscid fluid of constant density $\rho$.  The free-surface Euler equations (\cite{Whitham}) is 
our reference model:
\begin{align} \label{F1}
\begin{split}
& u_{t} + uu_{x} + vu_{y} = - \frac{p_x}{\rho}, \;\   \mbox{for} \;\  h(x+U_{0}t), < y <\tilde{\zeta}(x,t), \\
&  v_{t} + uv_{x} + vv_{y} = - \frac{p_y}{\rho} - g ,\;\  \mbox{for} \;\  h(x+U_{0}t), < y <\tilde{\zeta}(x,t), \\
&  u_{x} + v_{y} = 0, \;\  \mbox{for} \;\  h(x+U_{0}t) < y <\tilde{\zeta}(x,t), \\
& p=P(x+U_{0}t),  \;\  \mbox{at} \;\ y = \tilde{\zeta}(x,t), \\
& v = \tilde{\zeta}_{t} + u\tilde{\zeta}_{x},  \;\ \mbox{at} \;\ y = \tilde{\zeta}(x,t), \\
& v =U_{0}h_x+uh_{x}, \;\ \mbox{at} \;\ y = h(x+U_{0}t). \\
\end{split}
\end{align}
The free surface is denoted by  $\tilde{\zeta}(x,t)$, the bottom topography by $h$ and the forcing pressure distribution over the free surface by $P$. 
We consider either a submarine obstacle  traveling with 
speed $U_0$ along the bottom or a
pressure distribution traveling along the free surface, also with speed $U_0$. 
The fluid velocity is denoted  by $(u,v)$, the gravity acceleration by $g$, while the pressure within the fluid body by $p$.
All functions $(u, v,\tilde{\zeta}$, h and $p)$ in equations   (\ref{F1}) are considered to be smooth  and periodic in $x$ with period $L$.

We are in a flow regime having a given background current of constant shear. With this in mind, write the velocity field in the form
\begin{align} \label{F2}
\begin{split}
(u,v)=\nabla\tilde{\phi}+(ay,0),
\end{split}
\end{align}
where $\tilde{\phi}$ is the velocity potential of the irrotational component of the flow while $-a$ is the given constant vorticity value.  Substitute (\ref{F2}) in (\ref{F1})  to obtain
\begin{align*}
\begin{split}
& \Delta\tilde{\phi}= 0, \;\  \mbox{for} \;\ -h_{0}+ h(x+U_0t) < y <\tilde{\zeta}(x,t), \\
& (U_0-ah_{0})h_{x} + ahh_{x} + \tilde{\phi}_{x}h_{x} =\tilde{\phi}_{y}, \;\ \mbox{at} \;\ y = -h_{0}+ h(x+U_0t),\\
& \tilde{\zeta}_{t}+(a\tilde{\zeta}+\tilde{\phi}_{x})\tilde{\zeta}_{x}-\tilde{\phi}_{y}=0,
\;\ \mbox{at} \;\ y = \tilde{\zeta}(x,t), \\
& \tilde{\phi}_{t}+\frac{1}{2}(\tilde{\phi}_{x}^2+\tilde{\phi}_{y}^{2})+a\tilde{\zeta}\tilde{\phi}_{x} +\tilde{\zeta}- a\tilde{\psi}= - \frac{P(x+U_0t)}{\rho}, \;\ \mbox{at} \;\ y = \tilde{\zeta}(x,t),
\end{split}
\end{align*}
where $\tilde{\psi}$ is the harmonic conjugate of $\tilde{\phi}$. In a moving frame, given by 
$x\rightarrow x+U_0t$, denoting
\(
\overline{\zeta}(x,t)\equiv \tilde{\zeta}(x-U_0t,t), \;\  \overline{\phi}(x,y,t)\equiv \tilde{\phi}(x-U_0t,y,t),
\)
these equations read as 
\begin{align} \label{F4}
\begin{split}
& \Delta\overline{\phi}= 0 \;\  \mbox{for} \;\ -h_{0}+ h(x) < y <\overline{\zeta}(x,t), \\
& (U_0-ah_{0})h_{x} + ahh_{x} + \overline{\phi}_{x}h_{x} =\overline{\phi}_{y} \;\ \mbox{at} \;\ y = -h_{0}+ h(x), \\
& \overline{\zeta}_{t}+(U_0+a\overline{\zeta}+\overline{\phi}_{x})\overline{\zeta}_{x}-\overline{\phi}_{y}=0
\;\ \mbox{at} \;\ y = \overline{\zeta}(x,t), \\
& \overline{\phi}_{t}+\frac{1}{2}(\overline{\phi}_{x}^2+\overline{\phi}_{y}^{2})+(U_0+a\overline{\zeta})\overline{\phi}_{x} +\overline{\zeta}- a\overline{\psi}= - \frac{P(x)}{\rho} \;\ \mbox{at} \;\ y = \overline{\zeta}(x,t).
\end{split}
\end{align}
We have thus prescribed the velocity field to have a constant vorticity, through the given background flow,
and to satisfy the Neumann condition (now) around a fixed submarine
obstacle.   The surface pressure distribution and the bottom topography are stationary in this reference frame.

We have the following characteristic scales:  $h_{0}$ for distance, $(gh_0)^{1/2}$
as a reference long-wave speed, $(h_{0}/g)^{1/2}$ for time 
and $\rho gh_{0}$ as a characteristic pressure.  
Using these characteristic scales, and abusing of notation, we put (\ref{F4}) in a dimensionless  form:
\begin{align} \label{F55}
\begin{split}
& \Delta\overline{\phi}= 0, \;\  \mbox{for} \;\ -1+ h(x) < y <\overline{\zeta}(x,t), \\
& (F-\Omega)h_{x} + \Omega hh_{x} + \overline{\phi}_{x}h_{x} =\overline{\phi}_{y}, \;\ \mbox{at} \;\ y = -1+ h(x), \\
& \overline{\zeta}_{t}+(F+\Omega\overline{\zeta}+\overline{\phi}_{x})\overline{\zeta}_{x}-\overline{\phi}_{y}=0,
\;\ \mbox{at} \;\ y = \overline{\zeta}(x,t), \\
& \overline{\phi}_{t}+\frac{1}{2}(\overline{\phi}_{x}^2+\overline{\phi}_{y}^{2})+(F+\Omega\overline{\zeta})\overline{\phi}_{x} +\overline{\zeta}- \Omega\overline{\psi}= - P(x), \;\ \mbox{at} \;\ y = \overline{\zeta}(x,t),
\end{split}
\end{align}
where the Froude number is  $F ={U_0}/{(gh_0)^{1/2}}$ and $\Omega = {ah_0}/{(gh_0)^{1/2}}$.  All the variables in (\ref{F55}) are dimensionless.
The dimensionless vorticity is given by $-\Omega$.
In the linear regime equations (\ref{F55}) are of the form
\begin{align} \label{F5}
\begin{split}
& \Delta\overline{\phi}= 0, \;\  \mbox{for} \;\ -1+ h(x) < y <0, \\
& (F-\Omega)h_{x} + \Omega hh_{x} + \overline{\phi}_{x}h_{x} =\overline{\phi}_{y}, \;\ \mbox{at} \;\ y = -1+ h(x), \\
& \overline{\zeta}_{t}+F\overline{\phi}_{x}=\overline{\phi}_{y},
\;\ \mbox{at} \;\ y = 0, \\
& \overline{\phi}_{t}+F\overline{\phi}_{x} +\overline{\zeta}- \Omega\overline{\psi}= - P(x), \;\ \mbox{at} \;\ y = 0.
\end{split}
\end{align}
Since we will study linear waves, from now on each time we mention {\it the Euler equations} we mean system (\ref{F5}).

Either for linear or nonlinear waves, the particle trajectory beneath a surface wave is governed by  the dynamical system 
\begin{align} \label{F6}
\begin{split}
& \frac{dx}{dt}= \overline{\phi}_{x} +\Omega y+F , \\
& \frac{dy}{dt}=\overline{\phi}_{y}, \\
& x(0)=x_0, \;\ y(0)=y_0.
\end{split}
\end{align}
To compute its vector field one needs $\overline{\phi}_{x}$ and $\overline{\phi}_{y}$ in the bulk of the fluid. These are obtained from the Euler equations, which
in the present study is given by system (\ref{F5}). 
From the dispersion relation of system (\ref{F5}) we have that the linear wave speed $c$ is given by
\begin{equation}\label{eqC}
c = F - \frac{\Omega\tanh(k)}{2k} \pm \frac{ \sqrt{\Omega^2\tanh^2(k)  + 4k\tanh(k)}}{2k},
\end{equation}
where we will consider the mode with the positive sign of the square root. 
Suppose we have a traveling wave solution in the form $\overline{\phi}(x,y,t) = \overline{\phi}(x-ct,y,t)$. If $(x(t),y(t))$ represents a particle trajectory 
given by the dynamical system (\ref{F6})  then, in the wave's moving frame $ X = x - ct \mbox{ and } Y = y,$ the trajectories $(X(t),Y(t))$
satisfy the dynamical system in the form
\begin{equation}\label{F7}
\begin{array}{l }
\dfrac{dX}{dt} = \overline{\phi}_X(X,Y,t) + \Omega Y + F -c = \overline{\psi}_Y(X,Y,t)  + \Omega Y + F -c, \\  \noalign{\bigskip}
\dfrac{dY}{dt} =  \overline{\phi}_Y(X,Y,t) = -\overline{\psi}_X(X,Y,t).   \\  \noalign{\bigskip}
\end{array}
\end{equation}

A simple calculation shows that we cannot have waves excessively long while seeking for a Kelvin cat eye structure. For the cat eye structure to 
exist we need the presence of stagnation points, in the wave's moving frame.  In the  linear regime we have small amplitude waves and therefore
a weak velocity potential. Thus to leading order the flow is dominated by the background current, with $\overline{\phi}_{X},\overline{\phi}_{Y}\approx 0$, and
a linear dynamical system such as  
\begin{equation}\label{SistemaMovel}
\begin{array}{l }
\dfrac{dX}{dt} \approx   \Omega Y + F -c , \\  \noalign{\bigskip}
\dfrac{dY}{dt} \approx  0. \\  \noalign{\bigskip}
\end{array}
\end{equation}
Denote by $(X^\star,Y^\star)$ a candidate position for a stagnation point. This yields  the critical point 
$\Omega Y^\star =  c - F,$ which by substituting the wave speed (\ref{eqC}) gives
\begin{equation}\label{eqYstar}
Y^\star = -\frac{\tanh(k)}{2k} \pm \sqrt{ \frac{\tanh^2(k)}{4k^2}  + \frac{\tanh(k)}{\Omega^2k}}.
\end{equation}
In the long wave limit  with $(k\rightarrow 0)$ we have that 
\begin{equation}\label{formulaestag}
Y^\star = -\frac{1}{2} \pm \sqrt{  \frac{1}{4}  + \frac{1}{\Omega^2}}.
\end{equation}
No matter how strong is the vorticity of the background flow, we cannot find  $-1\leq Y^\star\leq0$. In other words there are no stagnation points within our
fluid body. For linear waves the stagnation point will appear as a balance between wavelength, total depth and the vorticity applied. 

In the next section we detail the conformal mapping technique for solving the Euler equations (\ref{F5})  and the dynamical systems  (\ref{F6}) or (\ref{F7}) in 
the canonical domain, where we have a flat strip.



\subsection{\bf  Conformal mapping}

\begin{figure}
	\centerline{\includegraphics[width=0.9\textwidth]{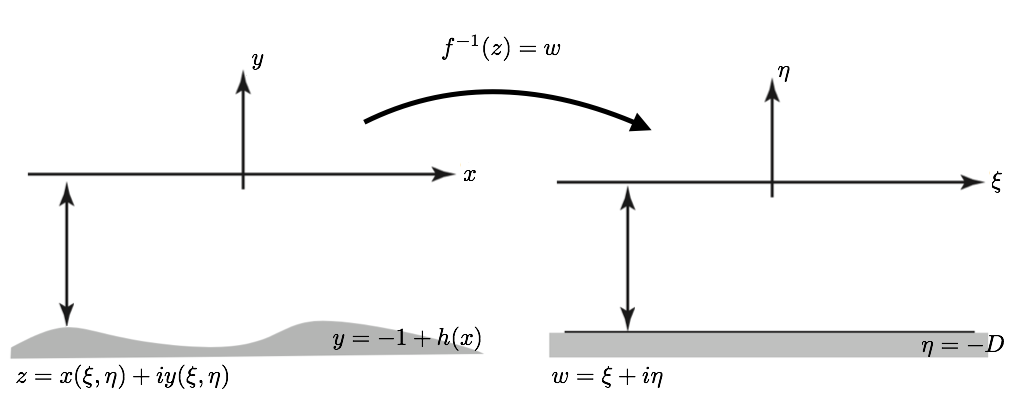}}
	\caption{The inverse conformal mapping. The bottom topography is flattened out in the canonical domain.}
	\label{fig:M.1}
\end{figure}

As depicted in figure \ref{fig:M.1}
the conformal mapping $z=f(w)$, from the canonical domain (a flat strip) onto the physical domain, is defined by  
\begin{equation*}
z(\xi,\eta) = x(\xi,\eta)+iy(\xi,\eta), 
\end{equation*}
where in the canonical $w$-plane $w=\xi+ i \eta$. We have the following boundary conditions: 
\begin{equation*}
y(\xi,0)=0 \;\ \mbox { and } \;\ y(\xi,-D)=-1+H(\xi).
\end{equation*} 
We are imposing that the canonical upper boundary $\eta=0$ is mapped onto the undisturbed free surface $y=0$. We denote by
$H(\xi)=h(x(\xi,-D))$  the topography representation in the $\xi$-variable, running along the bottom of the flat strip. The flat bottom boundary 
$\eta=-D$ in the canonical domain is mapped onto the topography profile, which in the physical domain reads as $h(x)$.
The constant $D$ will be determined by setting that the period in the canonical domain is the same of the fluid domain.

We denote by $\mathbf{X}(\xi)$ and  $\mathbf{Y}(\xi)$ the traces of the respective harmonic functions along  $\eta=0$, and by $\mathbf{X}_{b}(\xi)$ the respective
trace along the bottom $\eta=-D$. The $L$-periodic harmonic function $y$ satisfies
\begin{align*}
\begin{split}
& y_{\xi\xi}+y_{\eta\eta}=0, \;\ \mbox{in} \;\ -D < \eta < 0, \\
& y = 0, \;\  \;\ \mbox{at} \;\  \eta =0, \\
& y = -1+H(\xi), \;\  \;\ \mbox{at} \;\  \eta =-D. \\
\end{split}
\end{align*}
Using a Fourier transform $\F$ in the $\xi$-variable, we have that
\begin{align}\label{eqYCM}
\begin{split}
&y(\xi, \eta) = \F_{k\neq0}^{-1}  \left[  \frac{-\coth(kD)  \sinh(k\eta)\hat{H}}{\cosh(kD)}   \right]  +  \frac{\eta (1 -\hat{H}(0))}{D}, \\
\end{split}
\end{align}
where $k = (2\pi/L)j$, $j \in \mathbb{Z}$, and
$$ \F_{k}[g(\xi)]=\widehat{g}(k)=\dfrac{1}{L}\int_{-L/2}^{L/2}g(\xi)e^{-ik \xi}d\xi,$$
$$ \F^{-1}[\widehat{g}(k)](\xi)=g(\xi)=\sum_{j=-\infty}^\infty \widehat{g}(k)e^{{ i k \xi}}.$$
The Cauchy-Riemann equation $x_\xi = y_\eta$ yields 
\begin{equation}\label{eqXCM}
x(\xi, \eta)  =\F_{k\neq0}^{-1}  \left[  \frac{i\coth(kD)  \cosh(k\eta)\hat{H}}{\cosh(kD)}   \right]  +  \frac{ 1 -\hat{H}(0)}{D}\xi. 
\end{equation}
From (\ref{eqXCM}) we obtain
\begin{equation*}
\mathbf{X}_b(\xi) = x(\xi, -D) =\F_{k\neq0}^{-1}  \left[ i\coth(kD) \hat{H}    \right]  +  \frac{ 1 -\hat{H}(0)}{D}\xi.
\end{equation*}
Again by the Cauchy-Riemann equation $x_\xi = y_\eta$ and (\ref{eqYCM}), we obtain an alternative (and equivalent) expression:
\begin{equation}\label{eqXb}
\mathbf{X}_b(\xi) =  \frac{ 1 -\hat{H}(0)}{D}\xi +  \F^{-1}\left[ \frac{i \coth(kD)}{\cosh^2(kD)} \widehat{H} \right]  + \F^{-1}\left[ i \tanh(kD) \widehat{H} \right].
\end{equation}

We have made the choice that the periodic solutions have the same period  in the canonical and physical domains. Let $L$ and $\lambda$ be the respective periods,
so that 
\begin{equation*}
\mathbf{X}(\xi = L/2) - \mathbf{X}(\xi = -L/2) = \lambda
\end{equation*}
with
\begin{equation}\label{mediaX}
\big<\mathbf{X}_{\xi}\big> \equiv\frac{1}{L}\int_{-L/2}^{L/2}\mathbf{X}_{\xi}(\xi)d\xi = 1.
\end{equation}
From (\ref{eqXCM}) and  $\big<\mathbf{X}_{\xi}\big> = \widehat{\mathbf{X}}_{\xi}(0)$, it follows that
\begin{equation}
1=\frac{\lambda}{L}=\frac{1-\widehat{H}(0,t)}{D},
~~\mbox{where} ~~
\label{eqD}
D = 1- \big<H\big>. 
\end{equation}
This is the depth of the canonical channel.

The velocity potential $\phi$ is a harmonic function in both domains.  Changing variables in the bottom boundary condition  $(\ref{F5})_2$ yields 
the elliptic problem
\begin{align*}
\begin{split}
& \phi_{\xi\xi}+\phi_{\eta\eta}=0, \;\ \mbox{in} \;\ -D < \eta < 0, \\
& \phi = \mathbf{\Phi}(\xi,t), \;\ \;\ \mbox{at} \;\ \eta =0, \\
& \phi_{\eta} = (F- \Omega)H_{\xi}(\xi)+\Omega HH_{\xi}(\xi), \;\ \mbox{at}~  \eta =-D. \\
\end{split}
\end{align*}
The notation is such that  $ \phi(\xi,\eta,t) \equiv \overline{\phi} (x(\xi,\eta),y(\xi,\eta),t) $ and   $ \psi(\xi,\eta,t) 
\equiv \overline{\psi} (x(\xi,\eta),y(\xi,\eta),t) $ is the 
harmonic conjugate expressed in the canonical domain. Their traces along  $\eta=0$ are denoted by $\mathbf{\Phi}(\xi,t)$  and $\mathbf{\Psi}(\xi,t)$, respectively.
These time-dependent traces will be updated through the free surfaces conditions.
The harmonic conjugate $\psi$ satisfies
\begin{align*}
\begin{split}
& \psi_{\xi\xi}+\psi_{\eta\eta}=0, \;\ \mbox{in} \;\ -D < \eta < 0, \\
& \psi = \mathbf{\Psi}(\xi,t), \;\  \mbox{at} \;\ \eta =0, \\
& \psi = -(F- \Omega)H(\xi)-\frac{\Omega}{2} H^{2}(\xi)+Q, \;\  \mbox{at} \;\ \eta =-D,\\
\end{split}
\end{align*}
where $Q=Q(t)$. Solving these elliptic problems in Fourier space leads to
\begin{align}\label{eq_Pot}
\begin{split}
&\phi(\xi,\eta,t) =  \F^{-1}\left[   \frac{\cosh(k(\eta+D))}{\cosh(kD)} \widehat{\mathbf{\Phi}}(k,t)   +   \left(  \frac{i (F-\Omega) \widehat{H} + i\frac{\Omega}{2} \widehat{H^2}  }{\cosh(kD)} \right) \sinh(k\eta) \right], \\
& \psi(\xi,\eta,t) = \F^{-1}\left[  \left(  \widehat{\mathbf{\Psi}}(k,t) +  \frac{ (F-\Omega)\widehat{H} + \frac{\Omega}{2} \widehat{H^2} }{\cosh(kD)}   \right)\frac{\sinh(k(D+\eta))}{\sinh(kD)}    - \right.  \\
&-   \left. \left(  \frac{(F-\Omega) \widehat{H} + \frac{\Omega}{2} \widehat{H^2}  }{\cosh(kD)} \right) \cosh(k\eta)   \right] - \frac{Q(t)}{D} \eta.
\end{split}
\end{align}
In order to find $Q(t)$, we start with the Cauchy-Riemann equation  $\psi_\eta = \phi_\xi$ and  use the periodicity in $\xi$ to obtain
$$ \widehat{\psi_\eta}(k=0,\eta,t) =\widehat{\phi_\xi}(k=0,\eta,t) = \frac{1}{L}  \int_{-L/2}^{L/2} \phi_\xi(\xi,\eta,t) \, d \xi = 0.$$
From $(\ref{eq_Pot})_2$ we have 
\begin{equation*}
\begin{split} 
\psi_\eta(\xi, \eta,t) =  &\F^{-1}\left[  \left(  \widehat{\mathbf{\Psi}}(k,t) +  \frac{ (F-\Omega)\widehat{H} + \frac{\Omega}{2} \widehat{H^2} }{\cosh(kD)}   \right)\frac{k\cosh(k(D+\eta))}{\sinh(kD)}    - \right.  \\
&-   \left. \left(  \frac{(F-\Omega) \widehat{H} + \frac{\Omega}{2} \widehat{H^2}  }{\cosh(kD)} \right) k\sinh(k\eta)   \right] - \frac{Q}{D}.
\end{split}
\end{equation*}
Imposing  $\langle \psi_\eta \rangle = 0$ in the equation above yields
\begin{equation*}
\left( \widehat{\mathbf{\Psi}}(0,t) + (F-\Omega)\widehat{H}(0) +  \frac{\Omega}{2} \widehat{H^2}(0)   \right)\frac{1}{D}    -   \frac{Q}{D}=0,
\end{equation*}
which is rewritten as 
\begin{equation*}
Q(t) =   \widehat{\mathbf{\Psi}}(0,t) + (F-\Omega)\widehat{H}(0) +  \frac{\Omega}{2} \widehat{H^2}(0).
\end{equation*}
Using  $-\phi_{\eta}=\psi_{\xi}$ in  (\ref{eq_Pot}), and evaluating over  $\eta=0$, gives
\begin{equation}\label{eqPot1}
\begin{array}{l}
\mathbf{\Phi}_\xi(\xi,t) = \F^{-1} \left[  -i \coth(kD) \left(   \widehat{\mathbf{\Psi}_\xi} (k,t)  + \dfrac{(F-\Omega)\widehat{H_\xi} + \frac{\Omega}{2} \widehat{\partial_\xi H^2}}{\cosh(kD)} \right) \right].
\end{array}
\end{equation}


The linearized kinematic condition is obtained from (\ref{F5}) and written as
\begin{equation*}
\frac{D}{Dt}\big(\overline{\zeta}(x,t)-y\big) = 0, \;\ \mbox{where} \;\ \frac{D}{Dt}\equiv \partial_{t}+F\partial_{x}+\overline{\phi}_{y}\partial_{y}.
\end{equation*}
In the canonical domain the wave elevation  is given as $N(\xi,t)$, where $\overline{\zeta}(\mathbf{X}(\xi),t)=y(\xi,N(\xi,t))$. 
The conformal mapping's Jacobian, evaluated along the frees surface, is denoted as  
$J(\xi,0)= \mathbf{X}_{\xi}^2(\xi)\equiv M(\xi)^{2}$
(\cite{Terreno}).
The kinematic condition infers
a material curve where, using the change of  variables $\partial_{\xi} = M(\xi)\partial_{x}$  and $\partial_{\eta} = M(\xi)\partial_{y}$, we get 
\begin{equation*}
\frac{\mathcal{D}}{\mathcal{D}t}\big(N(\xi,t)-\eta\big) = 0, \;\ \mbox{where} \;\ \frac{\mathcal{D}}{\mathcal{D}t}\equiv \partial_{t}+\frac{F}{M(\xi)}\partial_{\xi}+\frac{\phi_{\eta}}{M(\xi)^2}\partial_{\eta}.
\end{equation*}
The kinematic condition and Bernoulli law are rewritten from  (\ref{F5}) to read as

\begin{align*}
\begin{split}
& N_{t}+\frac{F}{M(\xi)}N_{\xi}=\frac{{\phi}_{\eta}}{M(\xi)^{2}},
\;\ \mbox{at} \;\ \eta = 0, \\
& \mathbf{\Phi}_{t}+\frac{F}{M(\xi)} \mathbf{\Phi}_{\xi} +\overline{\zeta}- \Omega\mathbf{\Psi}= - P(\mathbf{X}), \;\ \mbox{at} \;\ \eta = 0,
\end{split}
\end{align*}     
in the canonical variables.

In a similar fashion  as presented in \cite{Terreno}, we can reflect our domain about $\eta=0$ and obtain an odd extension of  $y(\xi,\eta)$ on the strip $-D<\eta<D$.
In the physical domain the image is a reflected channel with the undisturbed free surface ($y=0$) as the axis of symmetry. This does not change any
of the presentation above. Actually the boundary condition $y(\xi,0)=0$, at the beginning of this section, is automatically satisfied due to the reflection.
Under this reflection,  $y(\xi,\eta)$ is an odd function in  $\eta$ that can be Taylor expanded as
$$ y(\xi,\eta) =  y_\eta(\xi,0)\eta +  \frac{y_{\eta\eta\eta}(\xi,0)}{6}\eta^3 + \cdots ,$$
which for small wave elevations becomes
%
$$ y(\xi,N(\xi,t)) \approx M(\xi)N(\xi,t).$$
In the canonical domain the kinematic condition and Bernoulli law are 
\begin{equation}\label{eqCinPot}
\begin{array}{l}
N_t =-  \dfrac{F}{M(\xi)} N_\xi - \dfrac{\mathbf{\Psi}_\xi}{M(\xi)^2}, \\  \noalign{\bigskip}
\mathbf{\Phi}_t  = - M(\xi) N -  \dfrac{F}{M(\xi)} \mathbf{\Phi}_\xi + \Omega \mathbf{\Psi} - P(\mathbf{X}(\xi)). \\
\end{array}
\end{equation}
We are in position to rewrite the 2D Euler equations only in the 
canonical variable $\xi$. The dependence on the $\eta$-variable is implicitly built-in through the harmonic extension performed by the (Hilbert-type) 
Fourier operator containing a $\coth(kD)$ as a multiplier. These operators are defined below. The 2D Euler system is recast in the form
\begin{align}\label{eqEulerC3}
\begin{split}
& \mathbf{X}_{\xi} =  1  + \C_{k\neq0} \left[ \F^{-1}\left[ \dfrac{\hat{H}_{\xi}}{\cosh(kD)}\right]\right] \\ 
& \mathbf{\Phi}_{\xi} = -\C \left[    {\mathbf{\Psi_\xi}} (k,t)  + \F^{-1}\left[ \dfrac{(F-\Omega)\widehat{H_\xi} + \frac{\Omega}{2} \widehat{\partial_\xi H^2}}{\cosh(kD)} \right] \right]\\
& N_{t} = -\frac{F}{M(\xi)}N_{\xi}-\frac{\mathbf{\Psi}_{\xi}}{M(\xi)^2}, \\
& \mathbf{\Phi}_{t} = - M(\xi)N  -\frac{F}{M(\xi)}\mathbf{\Phi}_{\xi}+ \Omega\mathbf{\Psi}- P(\mathbf{X}(\xi)),
\end{split}
\end{align}
where  
\begin{align}\label{eqEulerC4}
\begin{split}
& \mathrm{H}(\xi) = h(\mathbf{X}_{b}(\xi)), \\
& \mathbf{X}_{b}(\xi)=  \xi +  \C_{k\ne 0}\left[\F^{-1}\left[ \frac{\widehat{H}}{\cosh^2(kD)}\right]  \right] + \F^{-1}\left[ i \tanh(kD) \widehat{H} \right] .
\end{split}
\end{align}
The Fourier operators are: $$\mathcal{C}[\cdot]\equiv\F^{-1}\left[i~\coth(kD)\F_{k}[\cdot]\right ] \mbox{~~and~~}  \mathcal{C}_{k\ne 0}[\cdot]\equiv\F^{-1}[i~\coth(kD)\F_{k\ne 0}[\cdot]].$$





\section{Numerical Method}\label{metodo_numerico}

Our main goal in this study is to  exhibit (numerically) the spontaneous formation of critical layers and its robustness in overcoming 
a bottom topography.  In the stationary wave regime, particle pathlines coincide with streamlines (\cite{JFM17}). When the surface disturbance is not a 
traveling wave we cannot use the streamfunction to depict the pathlines.
Hence, in order to see the formation and respective dynamics of the Kelvin's cat eye structure we need to compute several particle
orbits. These tracers will allow for the visualization of the critical layer. Due to the presence of a bottom topography we
adopt a conformal mapping from the canonical (flat strip) domain onto the physical domain. All the dynamics is computed in 
the (simpler) canonical domain and then mapped onto the physical domain for a proper visualization of the pathlines. 

In order to  describe our numerical procedure  the Euler formulation, presented earlier, is now connected with the tracer dynamics.
Recall that  particle orbits $({x}(t), {y}(t))$ are governed by system (\ref{F6}). 
In the canonical domain the pre-image of a trajectory is given by  $(\xi(t), \eta(t))$, 
where  the mapping onto the physical domain yields $ (\tilde{x}(t), \tilde{y}(t))  =  (x(\xi(t), \eta(t)), y(\xi(t), \eta(t)))$.
The dynamical system for computing the 
particle pathlines in the canonical domain, is  
\begin{equation}\label{eqTrajetoriasCanonico}
\left\{
\begin{array}{l}
\dfrac{d\xi}{dt} (t) = \dfrac{\phi_\xi(\xi,\eta,t)}{J(\xi,\eta)} + \dfrac{(\Omega~ y(\xi,\eta) + F)~x_\xi(\xi,\eta)}{J(\xi,\eta)}, \\  \noalign{\bigskip}
\dfrac{d\eta}{dt} (t)  = \dfrac{\phi_\eta(\xi,\eta,t)}{J(\xi,\eta)} - \dfrac{(\Omega~ y(\xi,\eta) + F)~y_\xi(\xi,\eta)}{J(\xi,\eta)}, \\  \noalign{\bigskip}
\xi(0) = \xi_0, \;\ \eta(0)=\eta_0.\\
\end{array} 
\right.
\end{equation}
The Jacobian is expressed as $J(\xi,\eta)=x_{\xi}^2(\xi,\eta)+y_{\xi}^2(\xi,\eta)$. 
The particle is initially located at $(\xi_0,\eta_0)$, the pre-image of $(x_0, y_0)$.  
The physical pathline is obtained from
\begin{equation}\label{eqX}
\tilde{x}(t) =  x(\xi(t), \eta(t))  =\F_{k\neq0}^{-1}  \left[  \frac{i\coth(kD)  \cosh(k\eta)\hat{H}}{\cosh(kD)}   \right]  +  \frac{ 1 -\hat{H}(0)}{D}\xi,
\end{equation}
\begin{align}\label{eqY}
\begin{split}
&\tilde{y}(t) = y(\xi(t), \eta(t)) = \F_{k\neq0}^{-1}  \left[  \frac{-\coth(kD)  \sinh(k\eta)\hat{H}}{\cosh(kD)}   \right]  +  \frac{\eta (1 -\hat{H}(0))}{D}.\\
\end{split}
\end{align}
Recall that in the canonical variables the topography is  denoted by $H(\xi)$.
All Fourier transforms $\F$ are in the $\xi$-variable, computed numerically through an FFT (Fast Fourier Transform).
In computing the vector field of the dynamical system (\ref{eqTrajetoriasCanonico}) we use the potential  $\phi$  given by its Fourier representation
\begin{equation}
\label{eqPot}
\phi(\xi,\eta,t) =  \F^{-1}\left[   \frac{\cosh(k(\eta+D))}{\cosh(kD)} \widehat{\mathbf{\Phi}}(k,t)   +   \left(  \frac{i (F-\Omega) \widehat{H} + i\frac{\Omega}{2} \widehat{H^2}  }{\cosh(kD)} \right) \sinh(k\eta) \right] .
\end{equation}
The dynamical system is not autonomous. Therefore its vector field is updated through the free surface conditions:
\begin{equation}
\left\{
\begin{array}{l}
N_{t} = -\dfrac{F}{M(\xi)}N_{\xi}-\dfrac{\mathbf{\Psi}_{\xi}}{M(\xi)^2} \\  \noalign{\bigskip} 
\mathbf{\Phi}_{t} = - M(\xi)N  -\dfrac{F}{M(\xi)}\mathbf{\Phi}_{\xi}+ \Omega\mathbf{\Psi}- P(\mathbf{X}).\\ \noalign{\bigskip}
\end{array} 
\right.
\label{FS}
\end{equation}
In other words, we evolve the boundary data for the velocity potential as well as for the wave profile. 
Within the fluid body, updated values of $\phi$ and its derivatives are readily available from expression (\ref{eqPot}).
The boundary data of its harmonic conjugate is obtained from 
\begin{equation}
\begin{array}{l}
\widehat{\mathbf{\Psi}}(k,t) =  i \tanh(kD)\widehat{\mathbf{\Phi}}(k,t) -  \left(  \dfrac{(F-\Omega)\widehat{H} +\frac{\Omega}{2} 
	\widehat{H^2}}{\cosh(kD)} \right).\\ \noalign{\bigskip}
\end{array}
\end{equation}
The term $P(\mathbf{X})$ in the dynamic condition above is used only when a traveling pressure distribution is considered along the free surface. 
We use the fourth order Runge-Kutta method for numerically evolving systems 
(\ref{eqTrajetoriasCanonico}) and (\ref{FS}).




As mentioned all Fourier transforms are approximated by the FFT on a uniform grid, with all derivatives  performed in Fourier space (\cite{Trefethen,MilewskiTabak}). 
The computational grid in the canonical domain is given by $\xi\in[-L/2,L/2)$, with $N$ uniformly spaced  points, with grid size $\Delta \xi= L/N$. 
This corresponds to a nonuniform grid in physical space. A typical resolution has $N= 2^{13}$  with a time step $\Delta t = 0.05$.  
A Kelvin cat eye structure is typically captured with a cloud of $90$ tracers.

We have shown that   $\mathbf{X}_{b}(\xi)$ and  $H(\xi)$ are coupled in a nontrivial fashion. We need to beforehand compute the 
topography profile $H(\xi)$ in the canonical domain. This depends on a nontrivial composition of the form $h(x(\xi,-D))$. 
This topographic composition is pre-processed using an iterative method as presented by \cite{Marcelo-Paul-Andre}. The iterates are labelled by 
a superscript $l$.
The iterative scheme has the following structure:
\begin{align}\label{M7}
\begin{split} 
& \mathbf{X}^{l}_{b}(\xi)=  \xi +  \C_{k\ne 0}\left[\F^{-1}\left[ \frac{\widehat{H^{l}}}{\cosh^2(kD)}\right]  \right] + \F^{-1}\left[ i \tanh(kD) \widehat{H^{l}} \right] . \\
& H^{l+1}(\xi) = h(\mathbf{X}_{b}^{l}(\xi)),
\end{split}
\end{align}
where the initial step is  based on the identities $\mathbf{X}_{b}^{0}(\xi)=\xi$ and $H^{1}(\xi)=h(\xi)$.
The  stopping criteria used is 
\begin{equation*}
\displaystyle\max_{\xi\in[-L/2,L/2)}\big| H^{l+1}(\xi) - H^{l}(\xi)\big|<\varepsilon,
\end{equation*}
where $\varepsilon$ is a given tolerance. In our simulations, we used $\varepsilon = 10^{-12}.$

%
%
%
%

When we have a traveling wave along the free surface the pathlines are identical to the streamlines.  When this is not the case 
one can still compute the streamlines.  
The potential's harmonic conjugate $\psi$ is readily available from the Fourier representation
\begin{equation}
\begin{array} {r}
\psi(\xi,\eta,t) = \F^{-1}\left[  \left(  \widehat{\mathbf{\Psi}}(k,t) +  \dfrac{ (F-\Omega)\widehat{H} + \frac{\Omega}{2} \widehat{H^2} }
{\cosh(kD)}   \right)\dfrac{\sinh(k(D+\eta))}{\sinh(kD)}    - \right.  \\
-   \left. \left(  \dfrac{(F-\Omega) \widehat{H} + \dfrac{\Omega}{2} \widehat{H^2}  }{\cosh(kD)} \right) \cosh(k\eta)   \right] - \dfrac{Q(t)}{D} \eta.
\end{array}
\end{equation}
In the moving frame we have that 
\begin{equation}\label{refMovel}
X = x - ct \mbox{ and } Y = y,
\end{equation}
where $c$ is the wave speed  given by  (\ref{eqC}).
The streamfunction $\psi_T$ is obtained by putting $\overline{\psi} $ together with the shear flow in the form
\begin{equation}\label{PSI_T}
\psi_T(X,Y,t) := \overline{\psi}(X,Y,t)  + \frac{\Omega Y^2}{2} + (F-c)Y.
\end{equation}
At any instant of time the streamfunction can be evaluated on a uniform grid in the canonical domain and mapped onto points in the physical domain
in order to trace its level curves there.  Yet, at each given time, we denote $\psi_T(t)\equiv\psi_T(X,Y,t)$ and consider  $\|\psi_T(t)\|_1$
as  its 1-norm evaluated on the computational grid.

\section{Results} \label{simulacoes}

It is well known in the literature that stagnation points and  critical layers can be formed beneath a periodic traveling wave in the presence of vorticity
(\cite{teles&peregrine, vasan&oliveras,JFM17,philo}). There are different forms of finding the initial wave profile,
depending whether we have in mind an irrotational (\cite{ DCDS14}) or a rotational surface traveling wave (\cite{JFM17}).  See the references within these two
papers.

In the present study the waves are linear and dispersive,
so we do not have a traveling wave. 
Nevertheless for weak dispersion the propagating waves change shape very slowly. A right propagating disturbance is obtained by pre-processing our desired 
wave profile over a flat bottom and keeping the right going mode.
The initial
wave elevation is usually positioned at a reasonable distance from the bottom topography so that there the Jacobian is effectively 
equal to 1, and therefore $N(\xi,0)=N_0(\xi)\equiv\zeta(x,0)=\zeta_0(x)$.  Hence the wave elevation representation in the canonical and physical domains
are identical.

\begin{figure}
	\centering
	\includegraphics{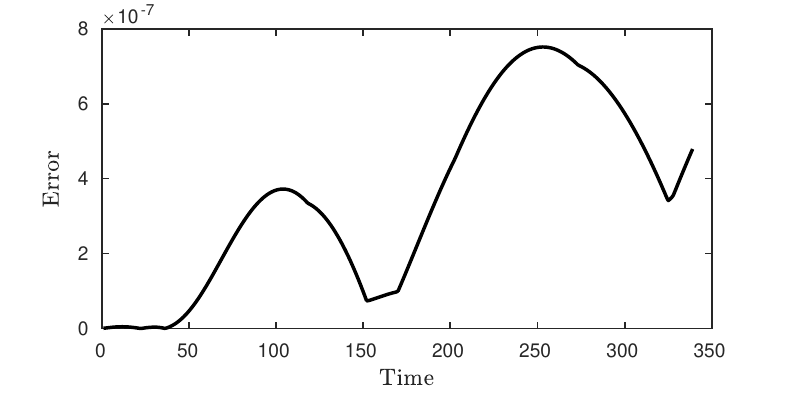}
	\caption{A nearly stationary stream function. The small relative error is given by  Error$(t)= {\| \psi_T(t) - \psi_T(0) \|_1}/{\|\psi_T(0)\|_1}$. 
	}
	\label{erroPsi}
\end{figure}

In the present study the periodic boundary condition is only for numerical purposes, namely for the Fourier spectral method. 
We consider wave disturbances localized in space  which  undergo reflection and transmission over a bottom variation of
compact support. Hence, over the time interval of our simulations, no disturbances should be observed at the endpoints of our 
computational domain. We consider a tabletop modulated wavetrain  localized in space which resembles a periodic wave
in its central region. The goal is to observe a segment of pathlines with a Kelvin's cat eye (critical layer) structure
which resembles  that of periodic waves.  

\subsection{Wave-current interaction for a slowly varying wavetrain.} \label{onda-correnteza}

\begin{figure}
	\centering
	\includegraphics[width=0.9\textwidth,height=6.cm]{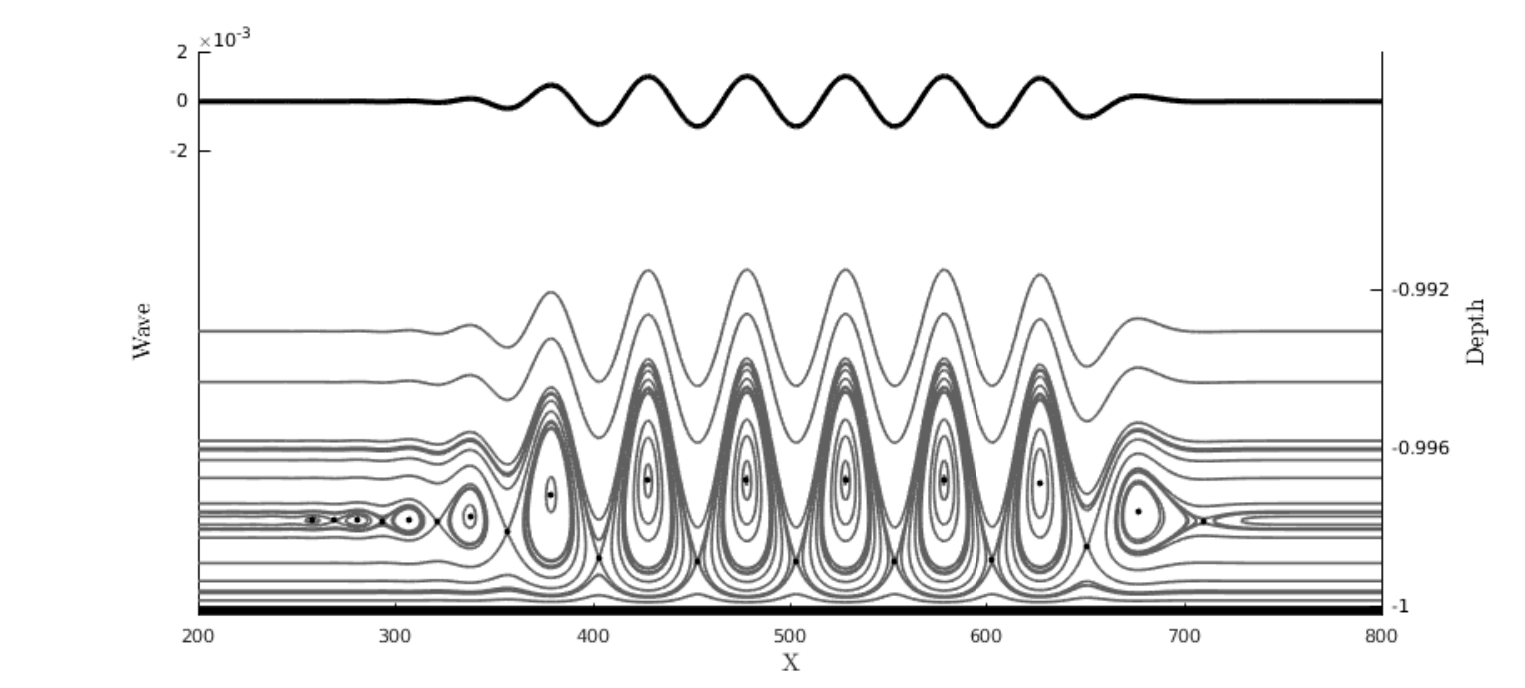}
	\caption{Streamlines obtained from $\psi_T(0)$. The dynamics is nearly stationary and the streamlines resemble  pathlines.  The vertical-axis to the left has the 
		scales corresponding to the wave elevation, namely the thicker line along the top.
		 The vertical-axis to the right   has values on the  depth scale and corresponds to the scales of the streamlines, indicating
		that the critical layer is very close to the bottom. 
	}
	\label{retrato1}
\end{figure}

We start with a tabletop wave profile of the form
\begin{equation} \label{dadosinciais}
N_0(\xi) =\alpha e^{-2(\sigma(\xi - \xi_0))^s} \cos(k(\xi-\xi_0)), 
\end{equation}
where $\alpha = 10^{-3}$, $\sigma = 6 \cdot 10^{-3}$, $s  = 8$ and $k =  2\pi/50$. The very first simulation is for an unforced case 
(with $h(x) = 0$ and $P(x) = 0$)  where the underlying current is such that  $\Omega = -18$ and $F = -9$. 
The dispersion is very weak in this case. The wave profile barely changes during the simulation.  As shown in figure \ref{erroPsi} the difference  
between the evolving profile and the initial disturbance is very small. The error 
was computed over a large time interval ($t\in[0,340]$) corresponding to a propagation distance of about 60 wavelengths. Over this time interval the 
level curves of the streamfunction are a good approximation for the pathlines.  The vector filed of equation    (\ref{F7}) is effectively autonomous. The 
initial phase portrait obtained from the streamfunction $\psi_T(0)$ is depicted in  figure \ref{retrato1}. 
The central region is typical of a phase portrait for a periodic traveling wave in the
presence of constant vorticity  (\cite{JFM17}). The novelty here is that the wavetrain is effectively of compact support, rather than periodic as in the previous studies. 
The super-Gaussian envelope, that provides a tabletop pattern over the central region  of the wavetrain, was designed so that we could observe 
a  Kelvin cat eye structure similar to periodic waves, as  presented by \cite{teles&peregrine}, \cite{vasan&oliveras} and \cite{JFM17} . 
This is the case in the region located approximately within the interval [400,600]. 
This is a first numerical display of the cat eye structure dynamically adjusting to the endpoints of the wavetrain. The wave is propagating
to the right and it is remarkable that beneath the very small (weakly dispersive) oscillatory  tail, the method can detect a number of 
diminishing cat eyes (recirculation regions) fading to the left. The small amplitude wave is linear and the Kelvin cat eye structure is narrow and located near 
the bottom. Nevertheless, due to a nontrivial vorticity (\cite{JFM17})  the cat eye structure is detached from the bottom exhibiting both types of critical points: saddles
and a center. The cat eye structure and associated critical layer is robust in the present of the weak dispersion. As the oscillatory tail of wave changes, 
the recirculation regions adjust accordingly. In particular, in analogy  with the development of an Airy-type solution, as the lagging oscillatory
tail develops, a small new recirculation (cat eye) region is generated.

\subsection{Critical layer robustness due to wave-current-topography interaction.} 

\begin{figure}
	\centering
	\includegraphics{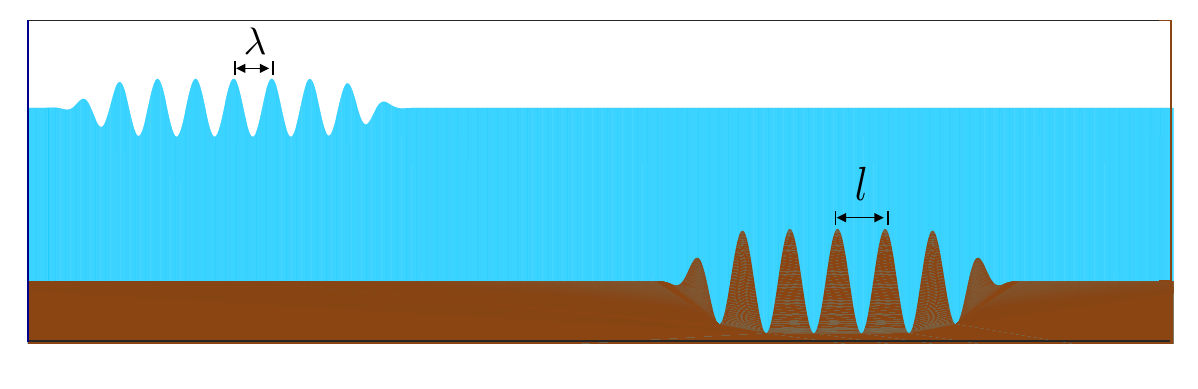}
	\caption{The physical domain is schematically depicted, for the wave-topography interaction, in the presence of a current of constant vorticity.  Two regimes are considered: 
		$\lambda \approx l$ (comparable wavelengths) and $\lambda \gg l$ (rapidly varying topography).}
	\label{dominio}
\end{figure}

Now we introduce a topography located in the middle of our computational domain. There is no pressure distribution along the free surface. 
The topography has compact support and 
a tabletop configuration, in order to locally resemble a periodic depth variation.  The bottom profile is given by

\begin{equation}
\label{topo}
h(x) =\delta e^{-2(\sigma(x - \tilde{x}_0))^s} \cos(k_b(x-\tilde{x}_0)),
\end{equation}
where $\delta\ll1$.  The right going initial wave profile is the same as before. 
We have kept the super-Gaussian (tabletop) modulation the same, for the wave and for the topography, so that
they have comparable lengths.
The physical
domain is schematically depicted in figure  \ref{dominio}. The initial disturbance has a Kelvin cat eye structure formed beneath it
and will eventually interact with the topography. At the same time a wave is generated by the current topography interaction, as studied in  \cite{Marcelo-Paul-Andre}. 
We will examine the submarine pathline structure in the presence of all these features.


\begin{figure}
	\centering
	\noindent\rule{0.9\textwidth}{0.4pt}
	\includegraphics[width=0.9\textwidth]{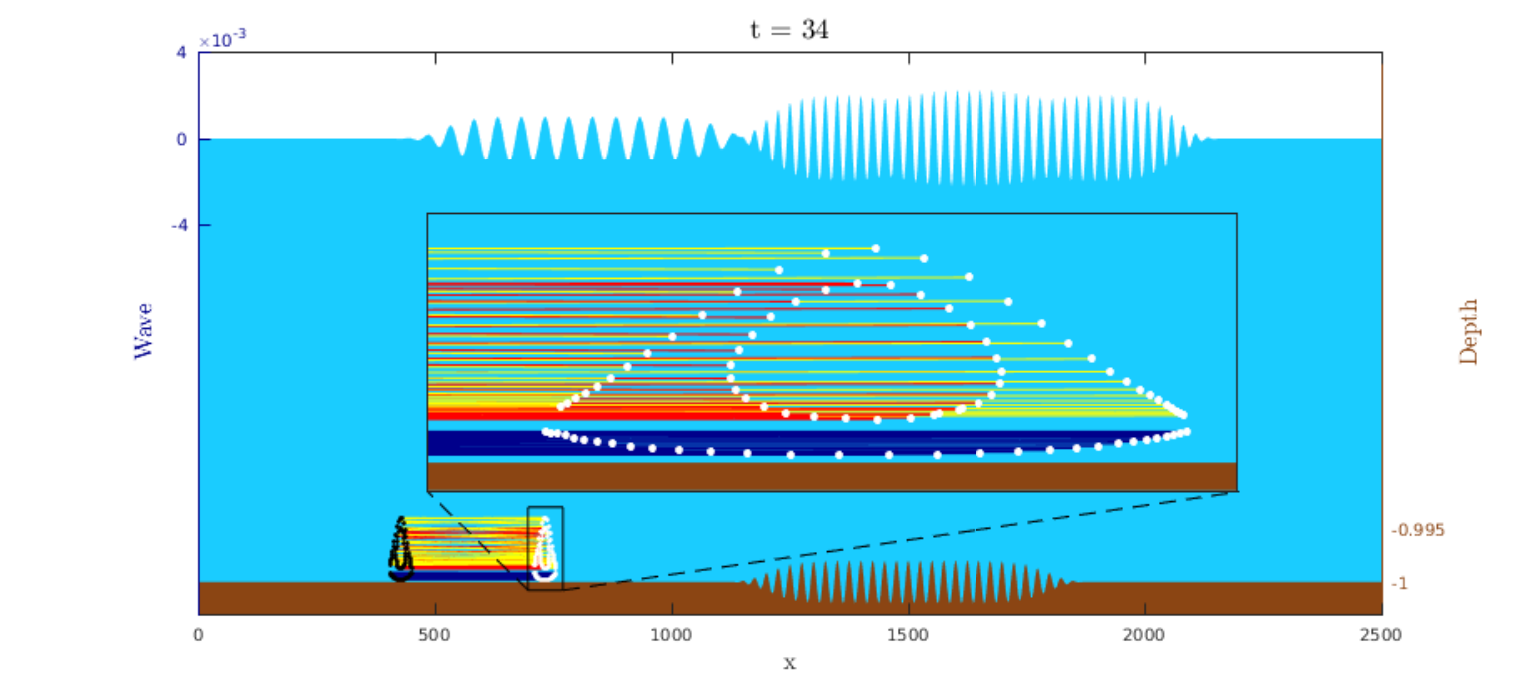}
	\noindent\rule{0.9\textwidth}{0.4pt}
	\includegraphics[width=0.9\textwidth]{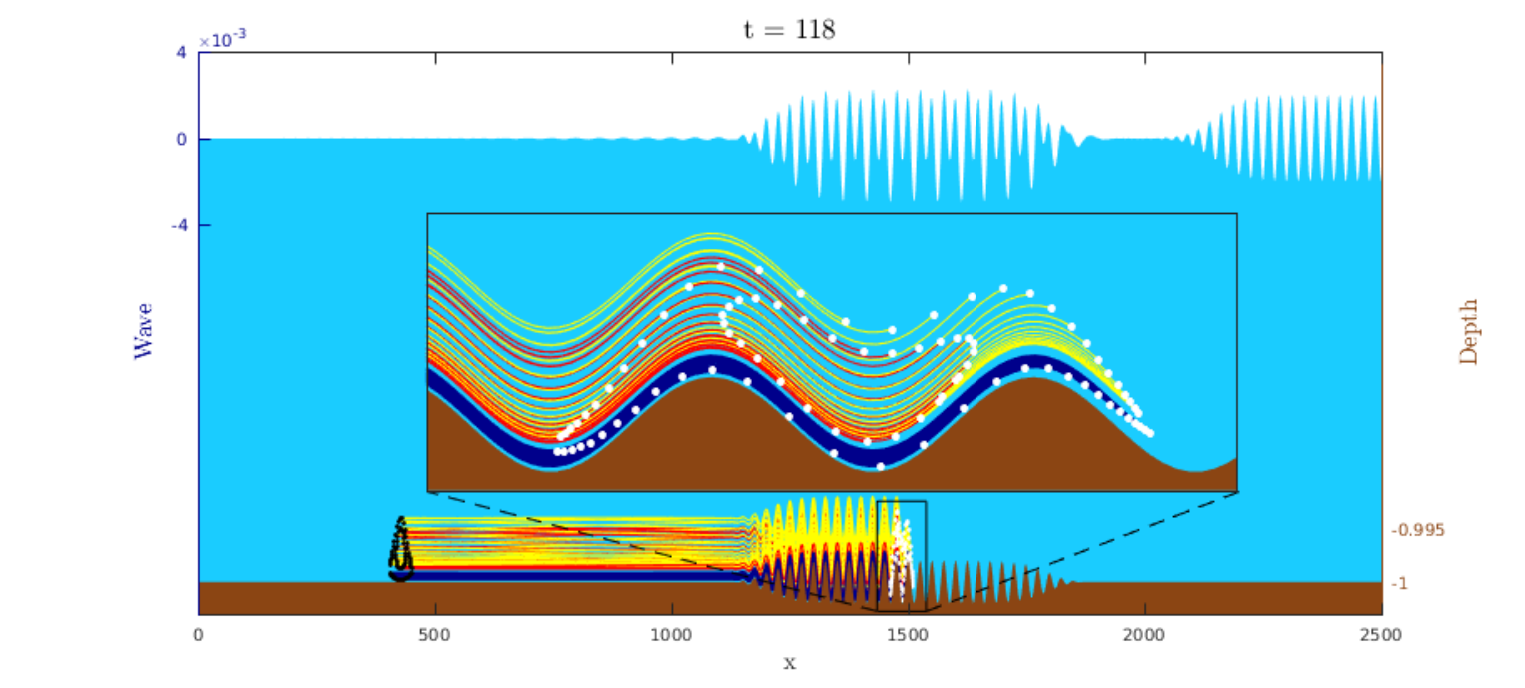}
	\noindent\rule{0.9\textwidth}{0.4pt}
	\includegraphics[width=0.9\textwidth]{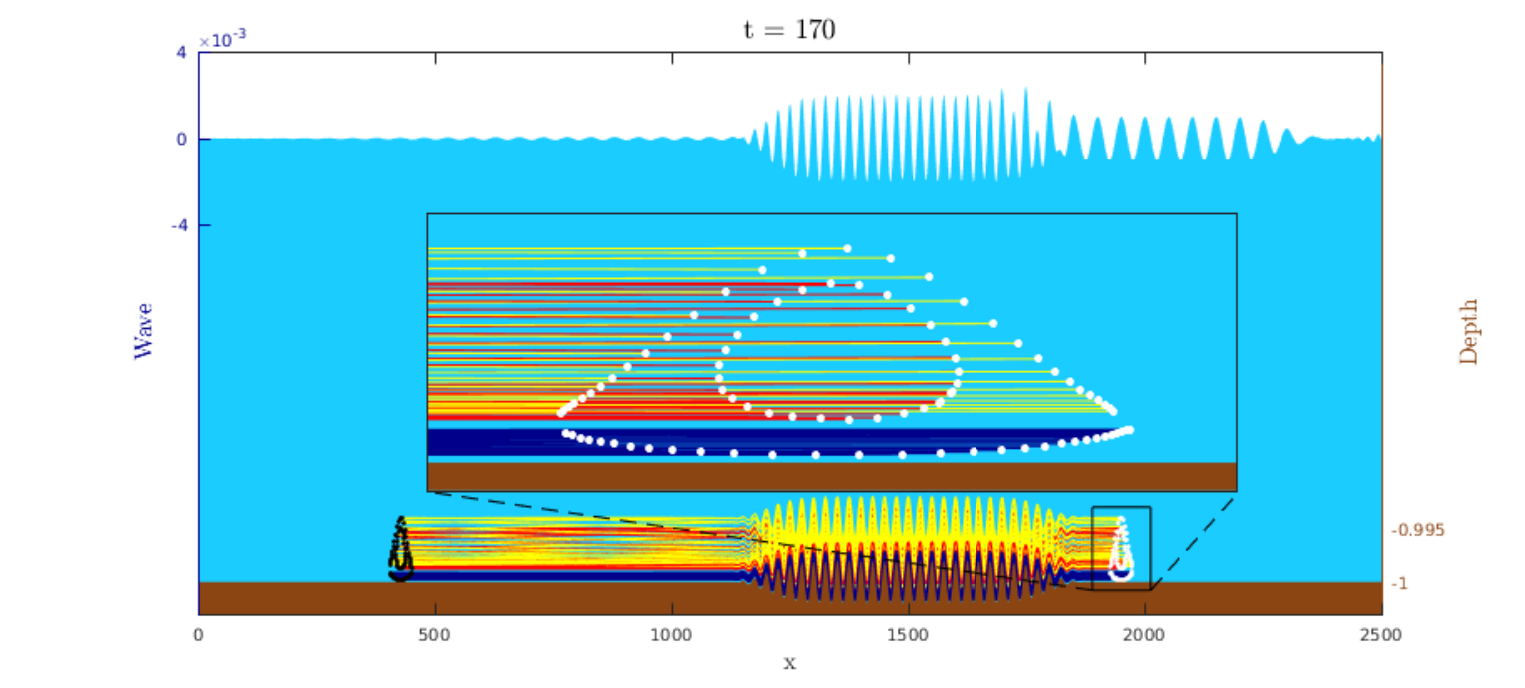}
	\caption{Comparable wavelengths regime ($\lambda =  2l$). The black dots indicate the initial position for each tracer, while the white dot 
		indicates their final position, at the end of the time interval. These are depicted in the {\bf laboratory (fixed) frame}. The $y$-axis to the left indicates the  scale
		for the free surface disturbance. The $y$-axis to the right indicates the depth values. The tracer and respective structures are very near the 
		bottom. Nevertheless our numerical method accurately captures the detailed structure of the tracers' orbits. 
		An animation (video01) can be found in the supplementary
material of this article.
			}
	\label{trajetoria_labframeBragg}
\end{figure}

We consider  two wave-topography regimes. First we consider the regime where the surface wavelength is comparable to that of the topography and we let 
$l/ \lambda=0.5$. Then we consider the case where the topography is rapidly varying: $ l/\lambda = 0.05$. The parameters used  for the underlying flow and the
initial disturbance (\ref{dadosinciais})
are:  $\Omega = -18$, $F = -9$,  $\alpha = 10^{-3}$, $\sigma = 3 \cdot 10^{-3}$, $s  = 8$,  $k =  2\pi/50$. 
For the topography we used $\delta = 2 \cdot 10^{-3}$,
$k_b = 2k$  and $k_b = 20k$ in expression (\ref{topo}).
The amplitude of the topography needs to be small so that the height of the waves generated by the current-topography interaction is comparable to that of the 
initial disturbance. This is clearly seen in the figures that follow.



Recall that in the present (non-stationary) regime the streamlines are not the same as the pathlines.  In order to visualize the pathlines we need to track the
orbits of many tracers, each tracer solving the dynamical system (\ref{eqTrajetoriasCanonico}).

In figure \ref{trajetoria_labframeBragg} we consider the $l/ \lambda=0.5$ case, first in the (fixed) laboratory frame. In the top snapshot (at time $t=34$) we 
see the initial (right going) disturbance at the left of the topography  and the current-generated wave above the topography. We focus on the initial 
disturbance and release a cloud tracers below this wave.  The initial cloud  (depicted by black particles near the bottom)  is 
positioned to the left of $x=500$.  At time $t=34$ the cloud has  traveled to the right, is positioned below the wave train, 
and is depicted by a bunch of white particles. The inset provides a  zoom of the white cloud and we clearly see the tracers in a Kelvin cat eye formation. 
The interaction of this structure with the topography is displayed in the middle picture of figure   \ref{trajetoria_labframeBragg}. 
At time $t=118$ we observe the pathlines in 
a terrain-following pattern, moving over the topography. The cat eye structure is distorted by the bottom undulations but retains its integrity. At this stage 
the rightgoing initial wave is interacting with the stationary 
wave generated by the current-topography mechanism. 
There is a nontrivial wave pattern above 
the topography, while a wave generated by the  current-topography  interaction is moving downstream and out of the picture. 
The computational domain is wider then that shown by these pictures.
The wave pattern due to the current-topography interaction is in accordance with that observed in  
\cite{Marcelo-Paul-Andre}, where there were standing waves above the topography together with waves moving downstream.
In the bottom picture at time $t=170$ the initial rightgoing disturbance is leaving the variable bottom region. The inset shows the tracers back to (nearly) the same
formation as at time $t=34$.
\begin{figure}
	\centering
	\noindent\rule{0.9\textwidth}{0.4pt}
	\includegraphics[width=0.9\textwidth]{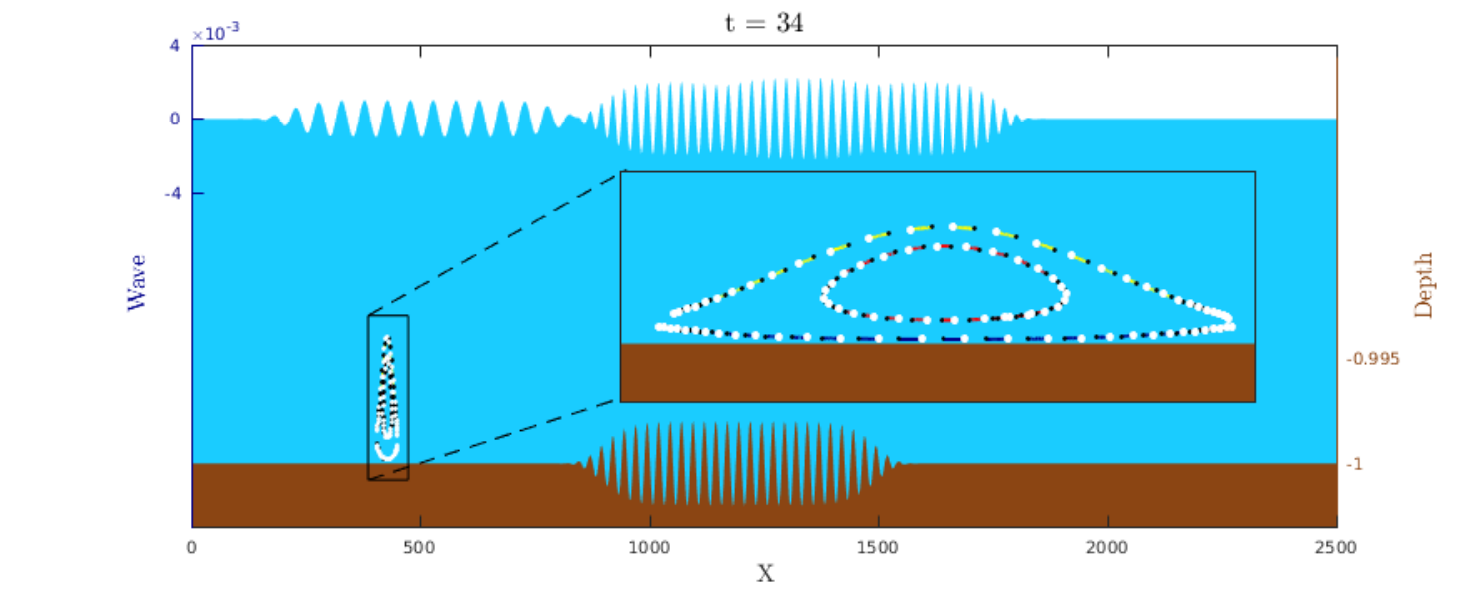}
	\noindent\rule{0.9\textwidth}{0.4pt}
	\includegraphics[width=0.9\textwidth]{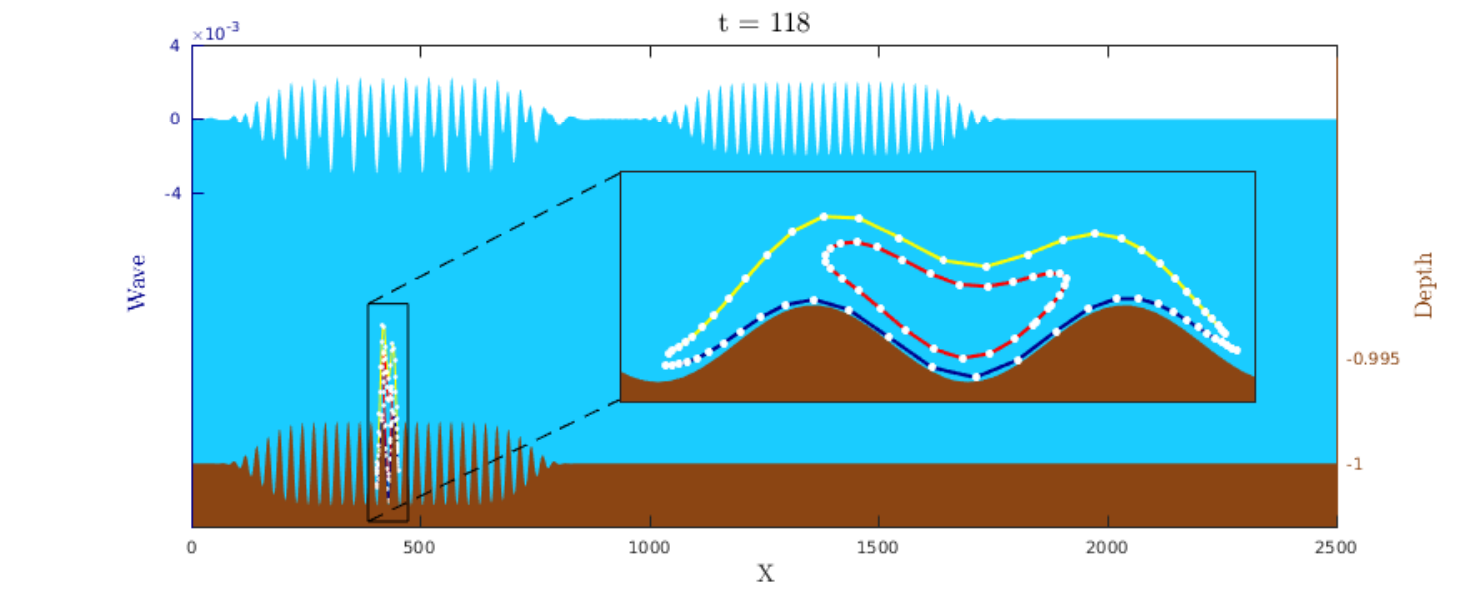}
	\noindent\rule{0.9\textwidth}{0.4pt}
	\includegraphics[width=0.9\textwidth]{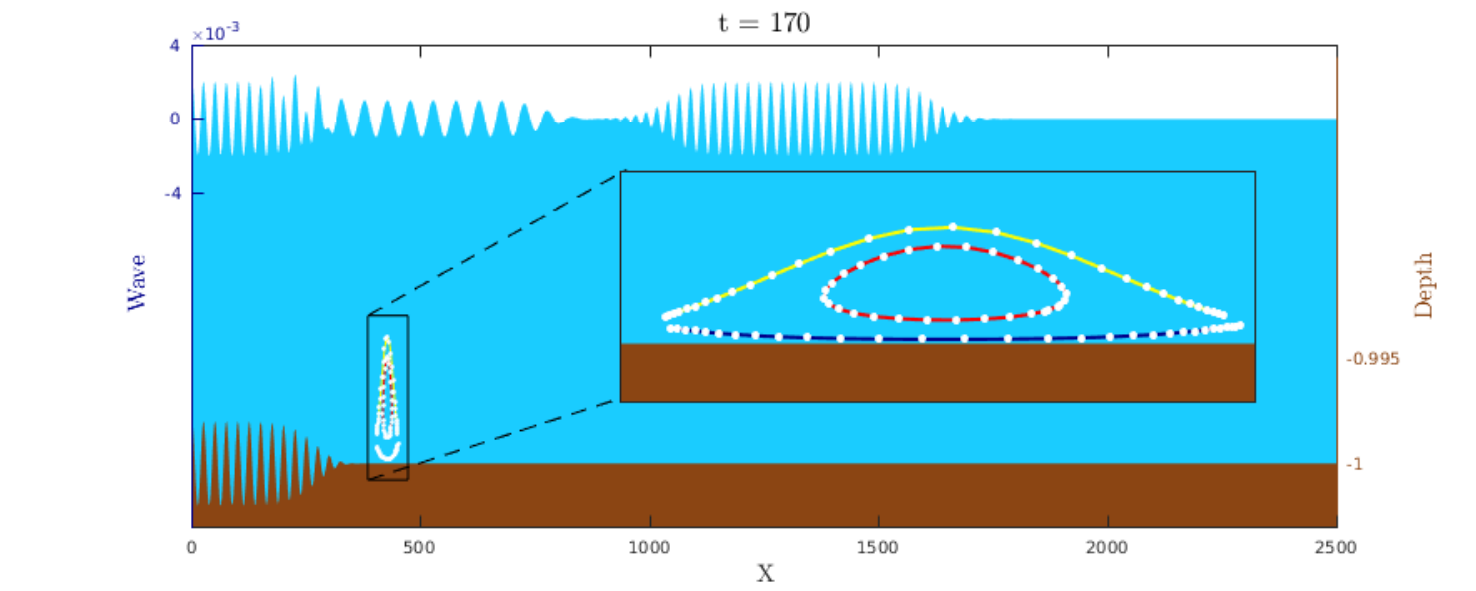}
	\caption{Comparable wavelengths regime ($\lambda = 2l$). The black dots indicate the initial position for each tracer, while the white dot 
		indicates their final position, at the end of the time interval. These are depicted in the {\bf wave's moving frame}. The $y$-axis to the left indicates the  scale
		for the free surface disturbance.  The $y$-axis to the right indicates the depth values near the bottom. The respective 
		cat eye structures are very small and very near the 
		bottom. Nevertheless the accuracy of the numerical method captures in great detail the Kelvin cat eye structure. 
		It is remarkable to observe the robustness of the recirculation region, adjusting to depth variations.
		An animation (video01) can be found in the supplementary
material of this article.
	}
	\label{trajetoria_MovframeBragg}
\end{figure}

%
\begin{figure}
	\centering
	\noindent\rule{0.9\textwidth}{0.4pt}
	\includegraphics[width=0.9\textwidth]{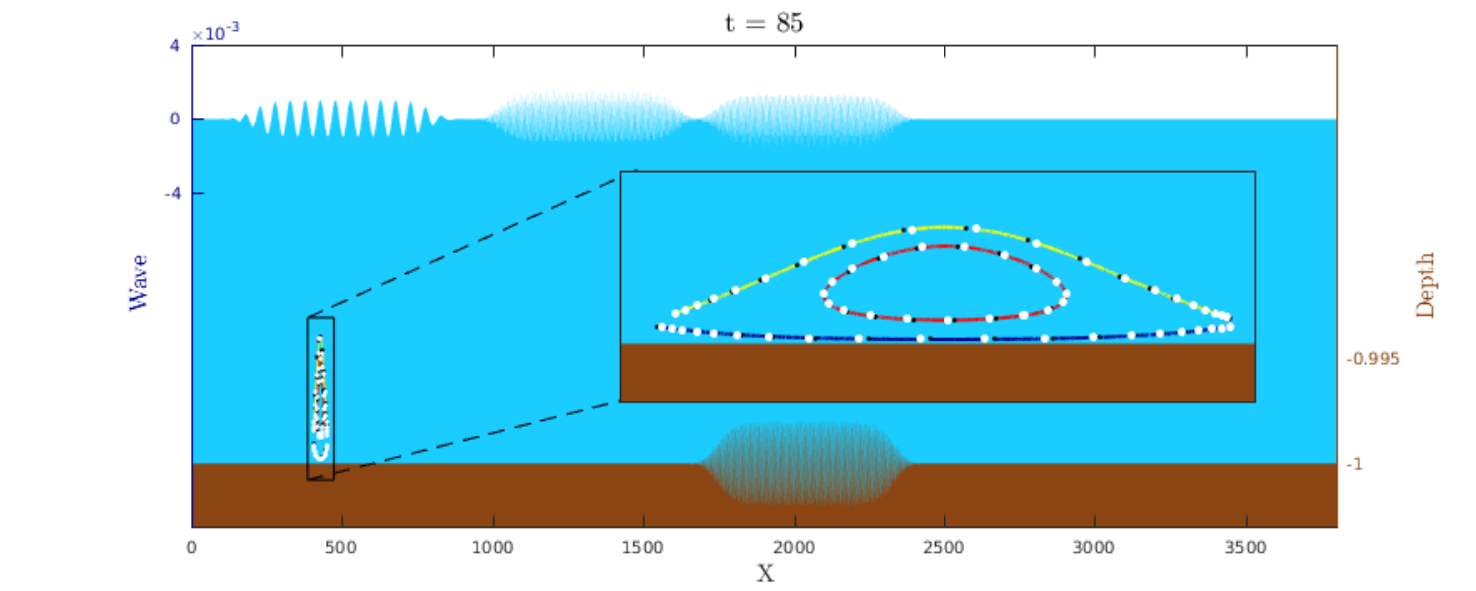}
	\noindent\rule{0.9\textwidth}{0.4pt}
	\includegraphics[width=0.9\textwidth]{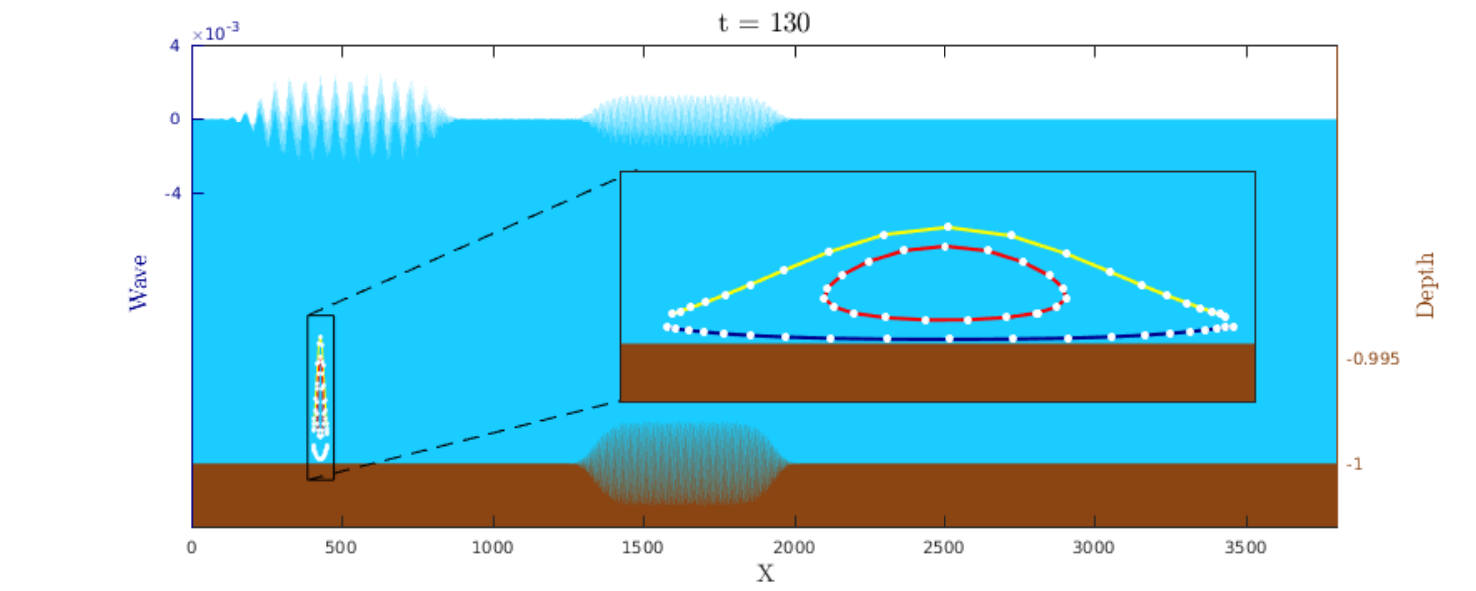}
	\noindent\rule{0.9\textwidth}{0.4pt}
	\includegraphics[width=0.9\textwidth]{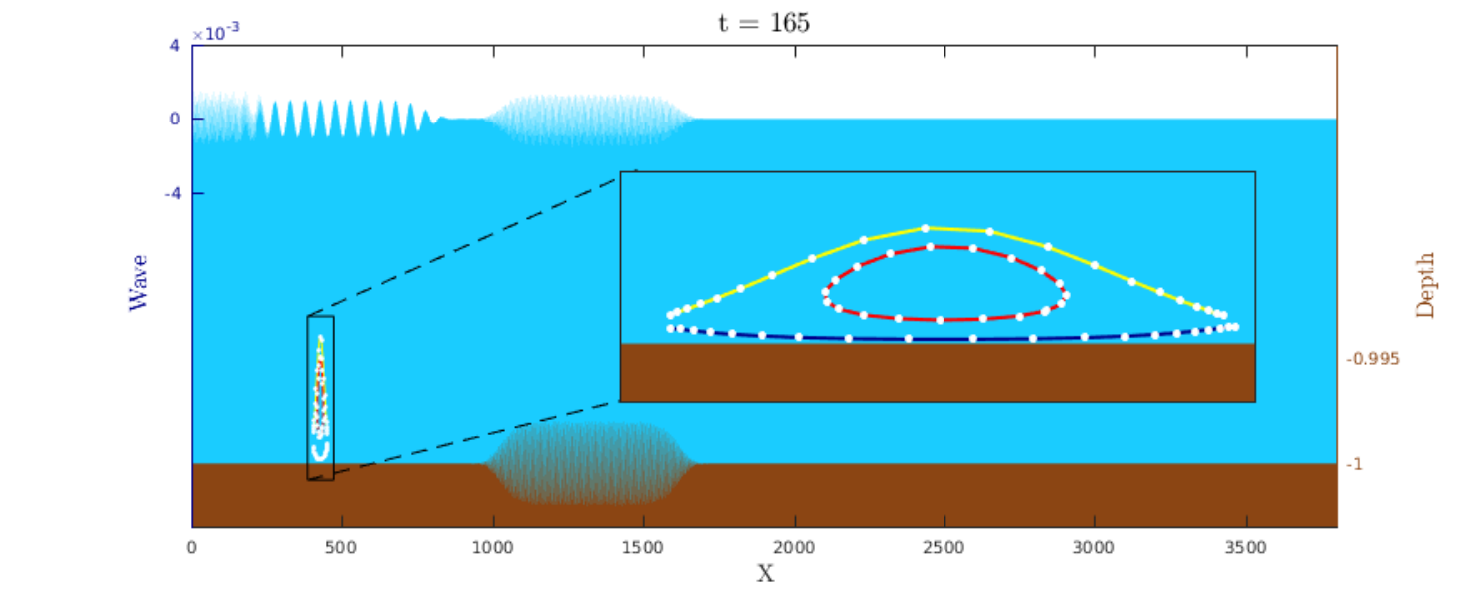}
	\caption{A rapidly varying topography ($\lambda \gg l$). We track tracers positioned beneath the initial rightgoing disturbance. 
		The black dots indicate the initial position for each tracer, while the white dot 
		indicates their final position, at the end of the time interval. These are depicted in the {\bf wave's moving frame}. For the middle and bottom picture we erased
		the initial position for a better visualization of the cat eye. The $y$-axis to the left indicates the  scale
		for the free surface disturbance.  The $y$-axis to the right indicates the depth values near the bottom. The respective 
		cat eye structures are very small and very near the 
		bottom. Nevertheless the accuracy of the numerical method captures in great detail the Kelvin cat eye formation. 
	}
	\label{FigEscalaRapida}
\end{figure}

The wave profile, due to the initial disturbance (\ref{dadosinciais}), changes very slowly. Hence we proceed as with traveling waves and display particle orbits in 
the  moving frame associated with the wave speed
$c$, given in (\ref{eqC}). These will reveal the Kelvin cat eye flow structure and the respective critical layer.
In figure \ref{trajetoria_MovframeBragg} the Kelvin cat eye structure is evident, with pathlines forming a closed recirculation region about a critical point
(centre), namely 
the stagnation point seen in \cite{JFM17}.  At the top
picture, the inset displays tracers that were initially located at the back points, but have  moved along the cat eye contours reaching the white points at time $t=34$. In the middle
picture, at time $t=118$, the inset displays material curves, where a collection of tracers is following a given contour of the cat eye structure. In the middle part (in red) we have a
closed material curve where the white points indicate the current position of each tracer. We have the same particles (cloud of tracers)  in all snapshots of figure
\ref{trajetoria_MovframeBragg}. The top material curve (in yellow) and the bottom material curve (in blue) eventually connect to the 
saddle points, as  identified by \cite{JFM17} and which were here 
displayed in figure \ref{retrato1}. The cat eye structure is very robust, as seen deforming to adjust to the topography while practically returning to its undisturbed shape
at time $t=170$, in the bottom picture of figure  \ref{trajetoria_MovframeBragg}.


%

We now consider a rapidly varying topography. The initial disturbance and collection of markers are the same from the previous example.
The current-topography interaction generates  rapidly varying wavetrains.  In the present regime there is a 
stationary wavetrain above the topography and another  wavetrain moving upstream. The initial rightgoing disturbance collides with the upstream wavetrain not
affecting at all the Kelvin cat eye, as shown in figure \ref{FigEscalaRapida}, at three different times.
As the wavetrain propagates over the rapidly varying topography the robustness feature is modified, in some sense.  In the top part of figure  \ref{FigEscalaRapida2}
we observe the disintegration of the material curves, which  formed the contours of the cat eye.  As soon as the wavetrain reaches the other
side of the depth variations, it is remarkable to observe that the same group of markers form again the same material curves of the cat eye (recirculation) region.

%


\begin{figure}
	\centering
	\noindent\rule{0.9\textwidth}{0.4pt}
	\includegraphics[width=0.9\textwidth]{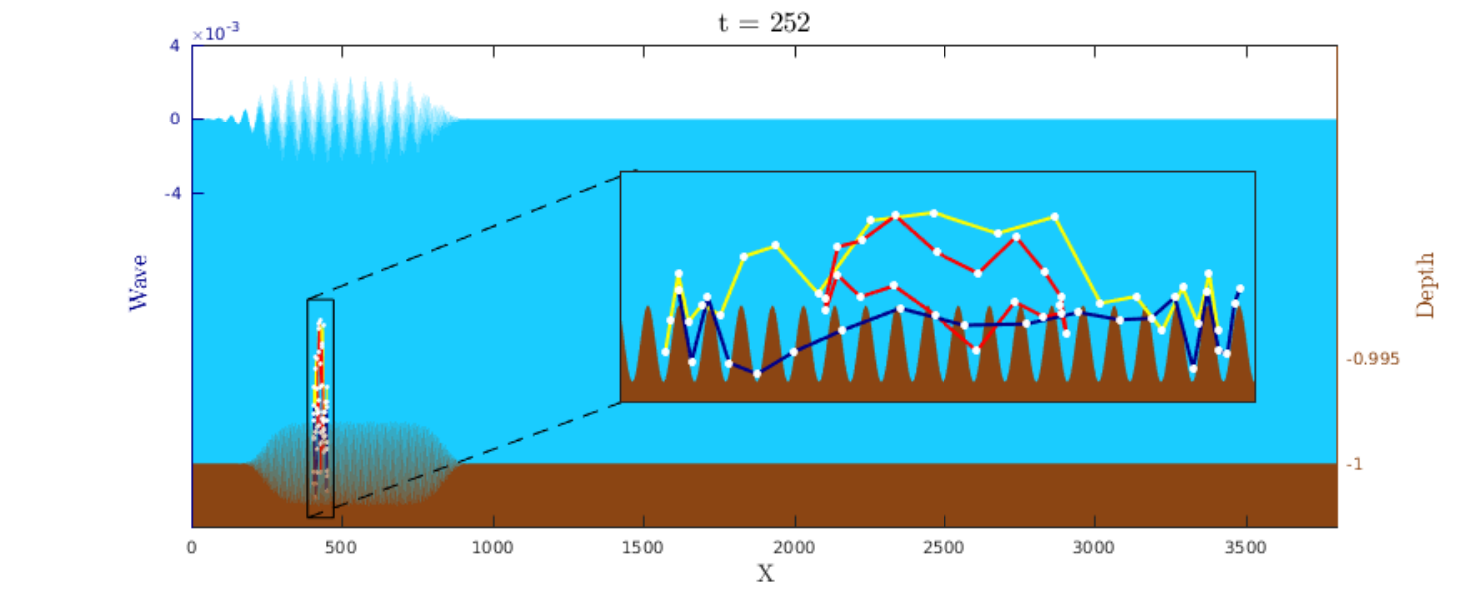}
	\noindent\rule{0.9\textwidth}{0.4pt}
	\includegraphics[width=0.9\textwidth]{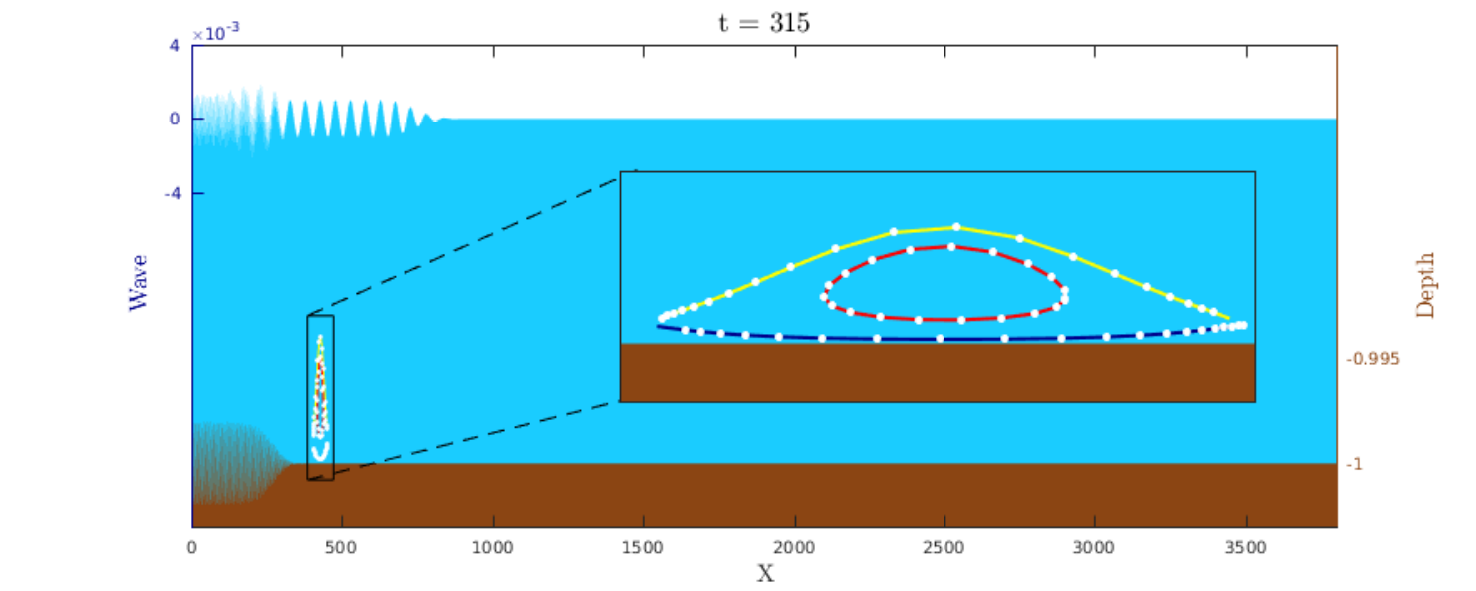}
	\caption{ Continuation of figure \ref{FigEscalaRapida}. As the wavetrain crosses over the rapidly varying topography, the bottom's fine features destroys
		the material curves' well defined formation, as seen at the top picture. Nevertheless as soon as the wavetrain leaves the variable depth region, the material curves 
		constituted with the 
		same particles as before, reconfigures itself in (namely) the same cat eye as observed in 
		figure \ref{FigEscalaRapida}. The robustness of the recirculation region, in this example, is
		remarkable. }
	\label{FigEscalaRapida2}
	
\end{figure}

\subsection{The spontaneous critical layer formation}

The spontaneous critical layer formation is associated with the spontaneous formation of stagnation points (in the moving frame), and the 
respective appearance of the Kelvin cat eye structure. Up to this point we have  tracers released beneath the initial wave disturbance. 
This initial wave disturbance immediately establishes the submarine phase-plane picture as seen in figure   \ref{retrato1}.
In the present case, the free surface starts at rest.
We then compute the submarine orbits for tracers placed beneath the wave generated by the 
current-topography interaction. At time $t=0^{+}$ there are no waves on the surface nor critical points in the bulk. The flow and topographic
parameters are the same as before, but  $k_b = k$. 

\begin{figure}
	\centering
	\noindent\rule{0.9\textwidth}{0.4pt}
	\includegraphics[width=0.9\textwidth]{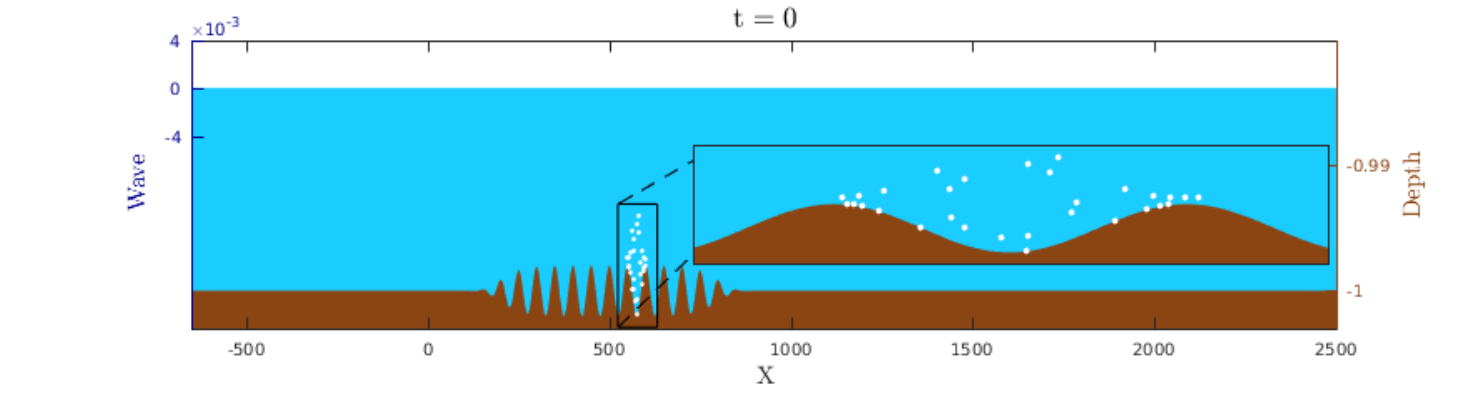}
	\noindent\rule{0.9\textwidth}{0.4pt}
	\includegraphics[width=0.9\textwidth]{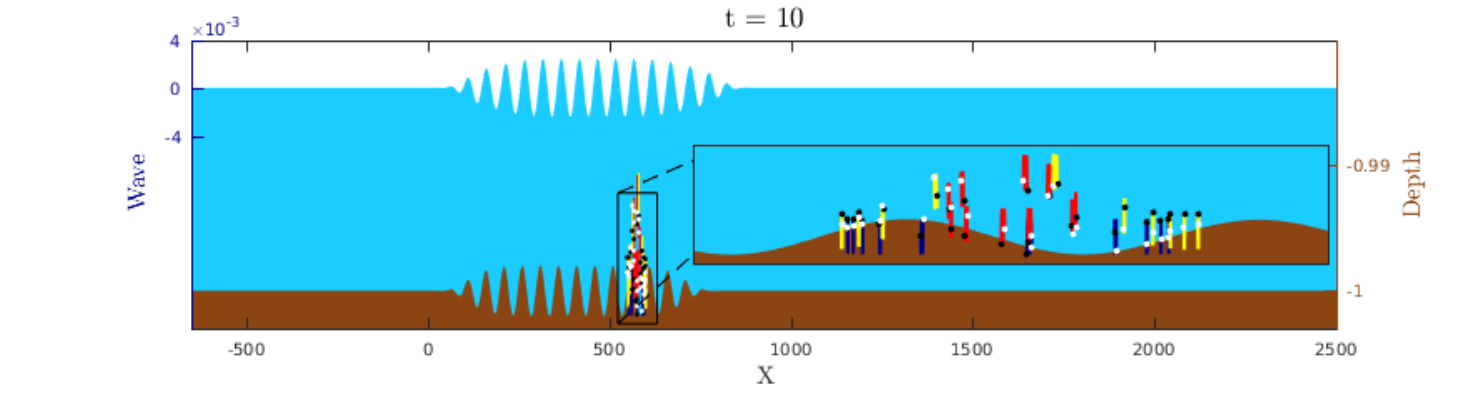}
	\noindent\rule{0.9\textwidth}{0.4pt}
	\includegraphics[width=0.9\textwidth]{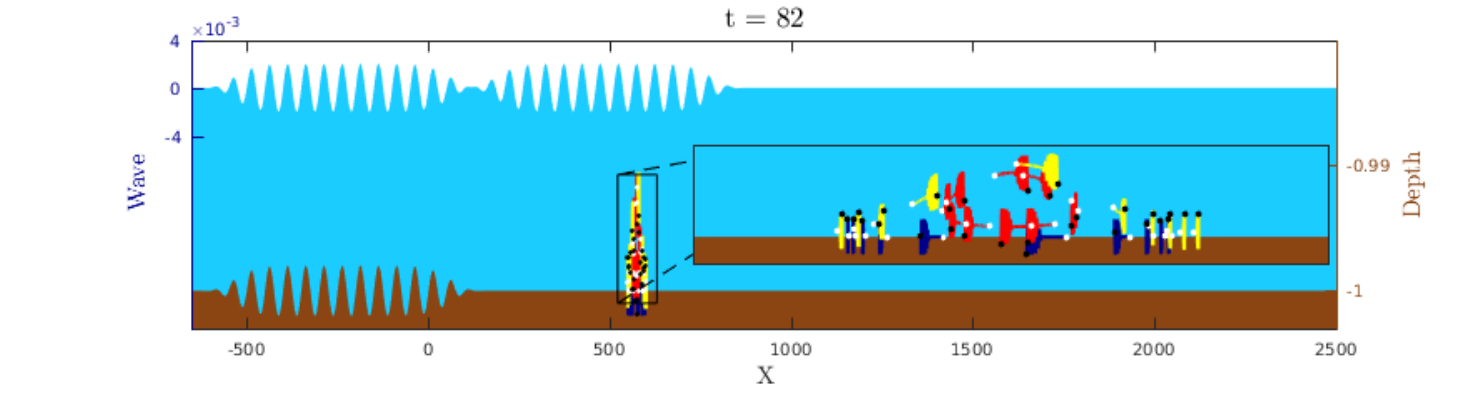}
	\caption{The spontaneous formation of the Kelvin cat eye structure. The top picture shows tracers positioned over a valley of the topography before the
		simulation starts. They are positioned in order to capture a cat eye formation. The middle picture shows  oscillating tracers  due to their relative 
		motion with the topography. 
		The black dot indicates the tracer's initial position while the white dot its final position at time $t=10$.  The bottom picture, at a later time ($t=82$), 
		shows a wavetrain generated by the topography, propagating downstream to the right. 
		The tracers' orbits 
		are depicted in the wave's moving frame, as we observe the topography moving to the left. Again the markers first oscillate vertically, after which we see the onset of the
		cat eye formation through the short orbit ``filament" connected to the final position (white dot). 
		The followup for this cat eye formation is presented in figure \ref{FigFundPer2}.
		The left $y$-axis has the wave scale while the right $y$-axis the depth scale.
		An animation (video02) can be found in the supplementary
material of this article.
		 }
	\label{FigFundPer}
\end{figure}


At time $t=0$ we distribute tracers at positions that would correspond to a cat eye. This is shown at the top picture
of figure \ref{FigFundPer}.  Then a standing wave appears, and at time $t=10$ we observe
the tracers moving up and down (middle picture of figure \ref{FigFundPer}). 
The color code (see version online) is the same as before, according to the position along the cat eye. Eventually a 
downstream propagating  wavetrain is generated by the topography.  
Note that the cloud of tracers is always centered about the same position because we are in the moving frame with respect to this wave. 
Therefore the topography seems to be moving to the left.
We  see at the bottom-picture of figure \ref{FigFundPer} that, after a 
preliminary oscillation, the markers start tracing  orbits (at $t=82$) which eventually will look like a cat eye. Erasing the initial oscillation, this is clearly seen  a bit later in 
time ($t=100$), at the top of figure \ref{FigFundPer2}. In the other two pictures of figure \ref{FigFundPer2}
the Kelvin cat eye is effectively a stationary structure. 

\begin{figure}
	\centering
	\noindent\rule{0.9\textwidth}{0.4pt}
	\includegraphics[width=0.9\textwidth]{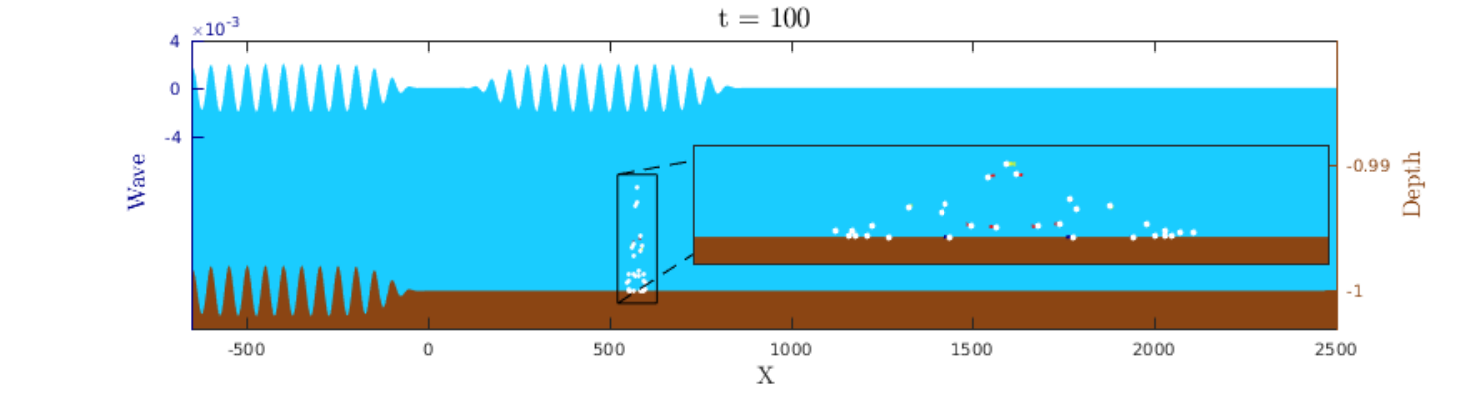}
	\noindent\rule{0.9\textwidth}{0.4pt}
	\includegraphics[width=0.9\textwidth]{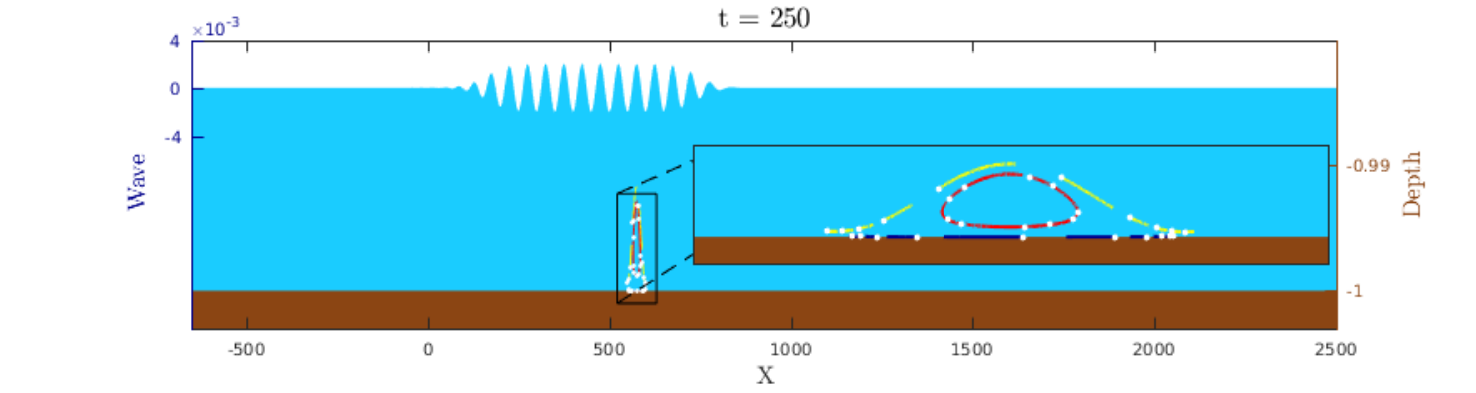}
	\noindent\rule{0.9\textwidth}{0.4pt}
	\includegraphics[width=0.9\textwidth]{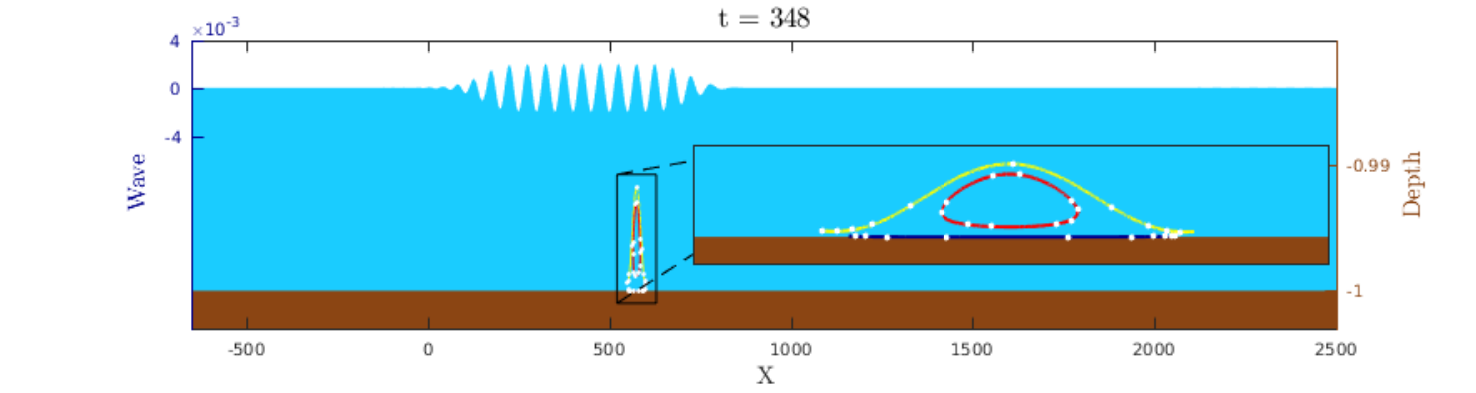}
	
	\caption{Continuation of figure \ref{FigFundPer}.  We have erased the initial positions of all tracers for a better visualization. As time evolves, from $t=100$ up to $t=348$, 
		we clearly see the Kelvin cat eye structure and  the associated critical layer. The overall picture for the pathlines is similar to that of figure \ref{retrato1}, 
		but a large amount of markers would be needed to clearly trace them. 
		An animation (video02) can be found in the supplementary
material of this article.
	}
	\label{FigFundPer2}
\end{figure}


The most common parameters used in numerical studies  for establishing the presence of critical layers under rotational waves are the vorticity, the energy and
the wavelength (\cite{ko&strauss2, ko&strauss1, philo, JFM17,teles&peregrine, vasan&oliveras}). In order to study the natural emergence of a critical 
layer, \cite{johnson} suggests a pressure distribution along the surface. 

%


\begin{figure}
	\centering
	\includegraphics[width=0.9\textwidth]{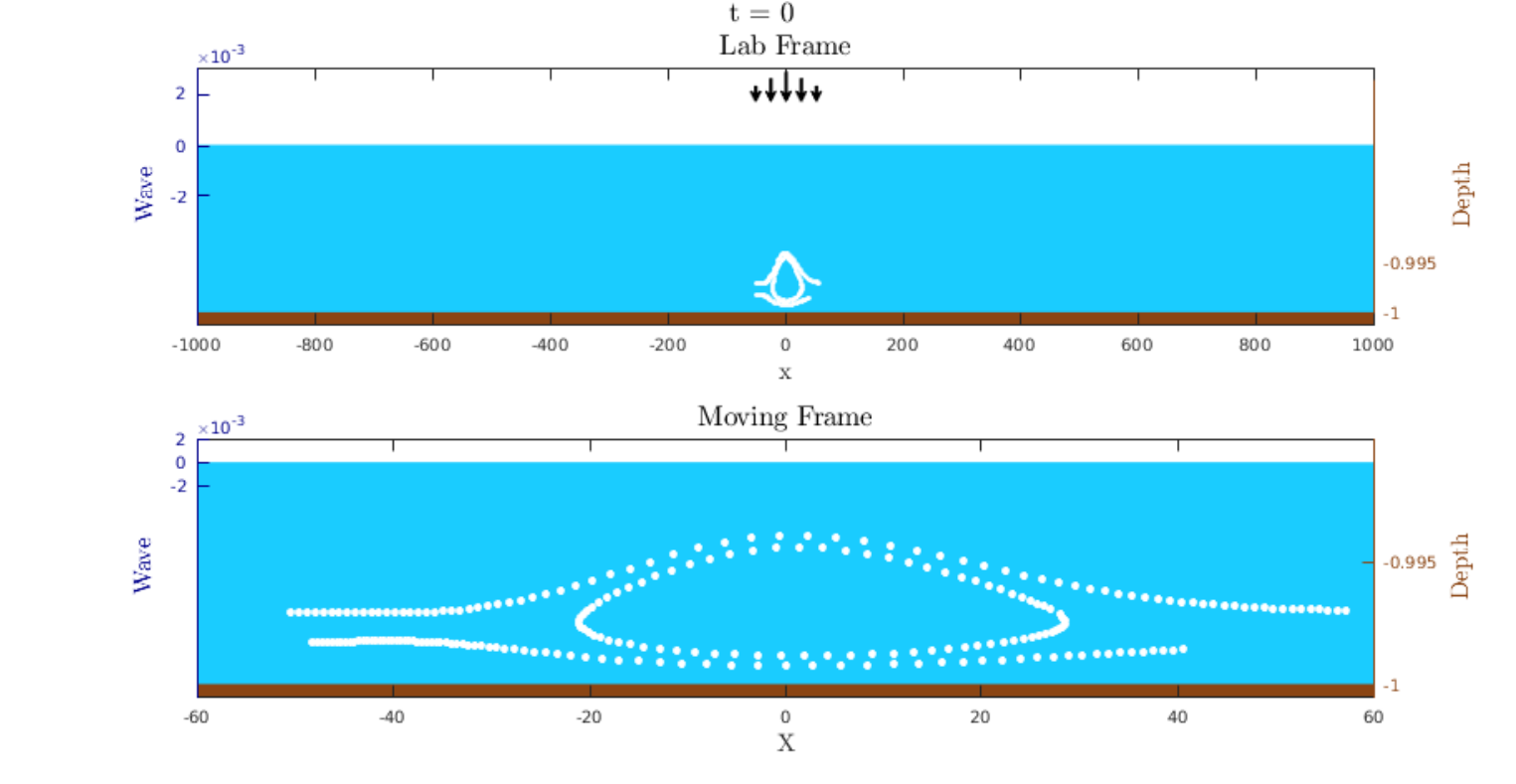}
	\noindent\rule{0.9\textwidth}{0.4pt}
	\includegraphics[width=0.9\textwidth]{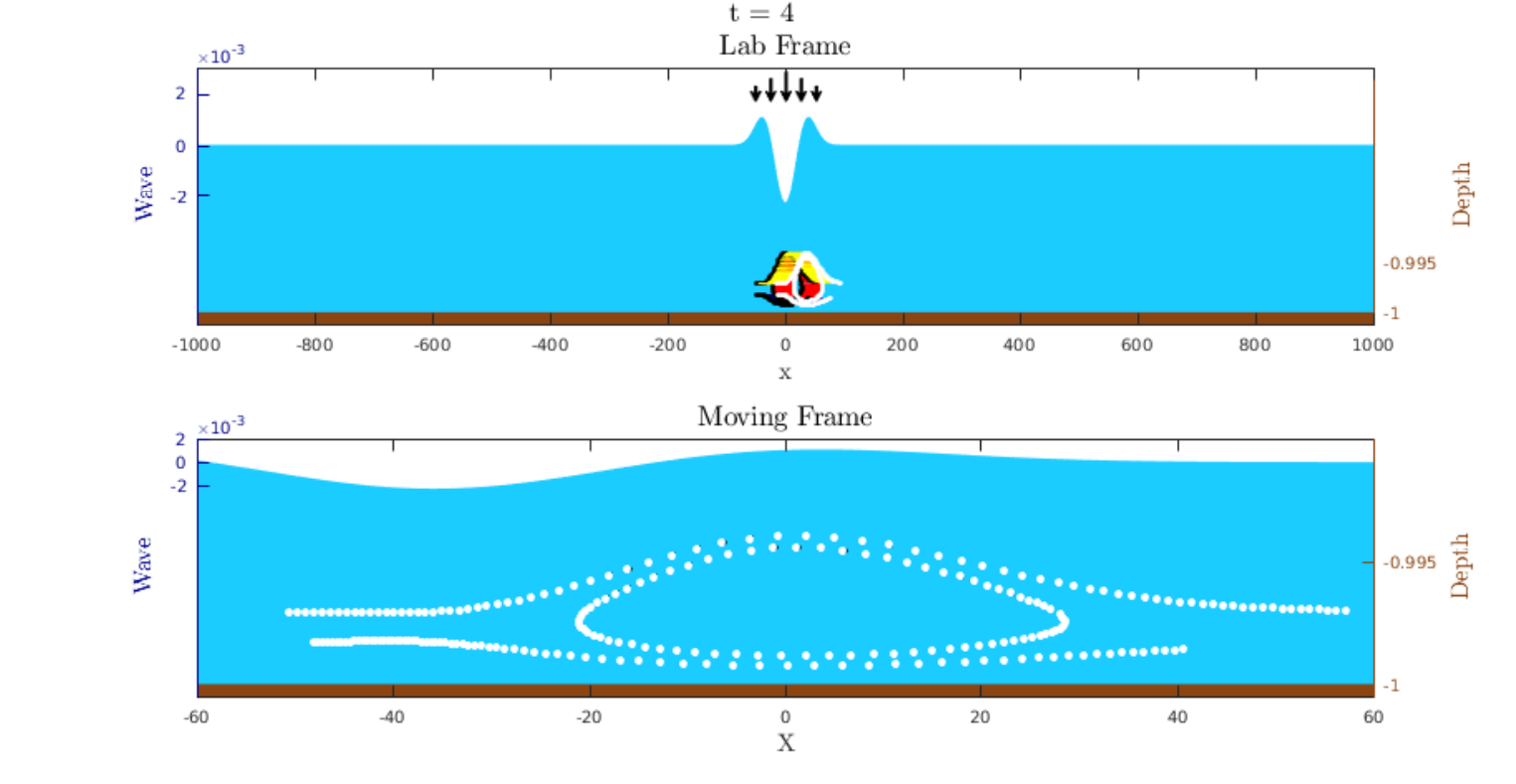}
	\caption{Wave generated by a localized pressure distribution. Top two pictures show the  markers initially positioned in order to capture the spontaneous cat eye 
		formation, both in the (fixed) laboratory frame as well as in the wave's moving frame. The bottom two pictures are at time $t=4$ at the onset of the pressure wave
		formation. A steady depression is observed where the pressure is applied while two disturbances are about to propagate outwards. The markers are set to follow the rightgoing
		pulse-shaped wave. In the laboratory frame we see a structure moving with the wave. In the moving frame the markers move very slowly. In figure  \ref{OndaPressao2}
		a large time interval is considered so that our cloud of markers has fully traced the cat eye structure. 
		The left $y$-axis has the wave scale while the right $y$-axis the depth scale.
		An animation (video03) can be found in the supplementary material of this article.
	}
	\label{OndaPressao}
\end{figure}

With this in mind, at time $t>0$,  we apply a localized (bell-shaped) high pressure distribution  $$P(x) = 0.2e^{-\sigma x^2}, $$ along the free surface. The bottom is flat 
($h(x) = 0$) and the free surface initially undisturbed ($ N_0(\xi) = 0, ~\mathbf{\Phi}(\xi,0) = 0)$. The underlying shear flow is such that $\Omega = -18$ and $F = -9$.


In figures  \ref{OndaPressao} and \ref{OndaPressao2} we display different stages of the submarine dynamics, both in the laboratory frame as well as in the 
wave's moving frame. 
The corresponding wave speed was found by inspection.
The top two pictures are at $t=0$, when the bell-shaped pressure is about to be applied and a cloud of markers has been prepared in the fluid 
body. At time $t=4$ the steady pressure distribution has generated two pulse-shaped waves which are about
to propagate to the left and right  accordingly. The tracers' dynamical systems are set to follow the rightgoing wave, as depicted by the two bottom pictures. The black dot
is the starting point while the white dot the final position for the respective time interval.  In the bottom picture of figure \ref{OndaPressao} these are almost at the 
same position because the dynamics is relative to the wave position. Hence in the moving frame each tracer position is slowly varying. The steady high pressure
distribution creates a steady free surface depression.

\begin{figure}
	\centering
	\includegraphics{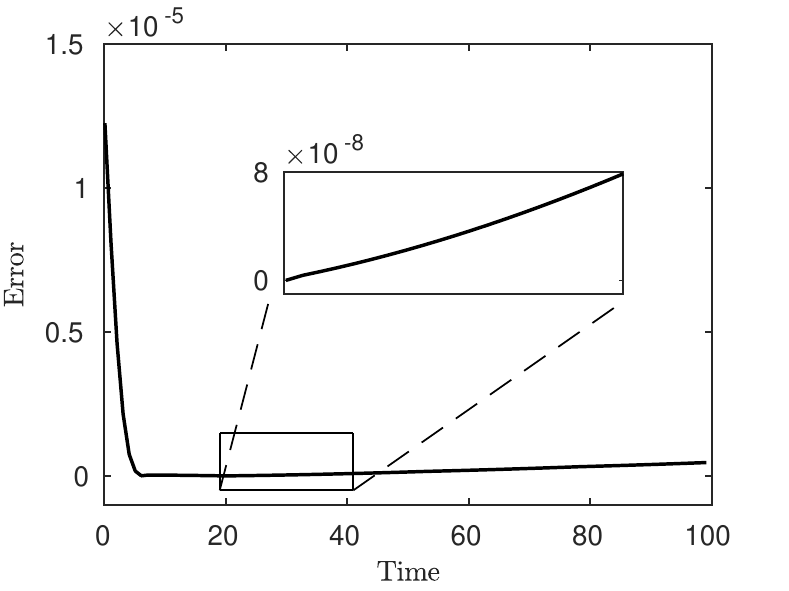}
	\caption{Effectively-steady streamfunction $\psi_T$. The small relative error, during the time interval $t \in [20,40]$, is computed as
		Error$(t) = {\| \psi_T(t) - \psi_T(20) \|_1}/{\|\psi_T(20)\|_1}$.  
		During an initial small time interval the stream function is quite different from its reference state $\psi_T(20)$. }
	\label{erroPsiPressao}
\end{figure}

Note that the waves generated are weakly dispersive. They are similar to a traveling  pulse-shaped wave for a certain time interval as seen at the
top of figure  \ref{OndaPressao2}. 
At time $t=99$ we observe a small dispersive tail appearing behind the leading pulse.   Nevertheless
the cat eye structure, spontaneously formed, remained unchanged as the wave propagated for more than 10 pulse widths in distance.
As shown figure   \ref{erroPsiPressao} the streamfunction  is effectively steady  during the time 
interval $t \in [20,40]$ where, up to a relative error of $O(10^{-8})$, the initial 
streamfuntion remains (approximately) unchanged. During this period of time the streamlines effectively coincide with the pathlines. 

\begin{figure}
	\centering
	\includegraphics[width=0.9\textwidth]{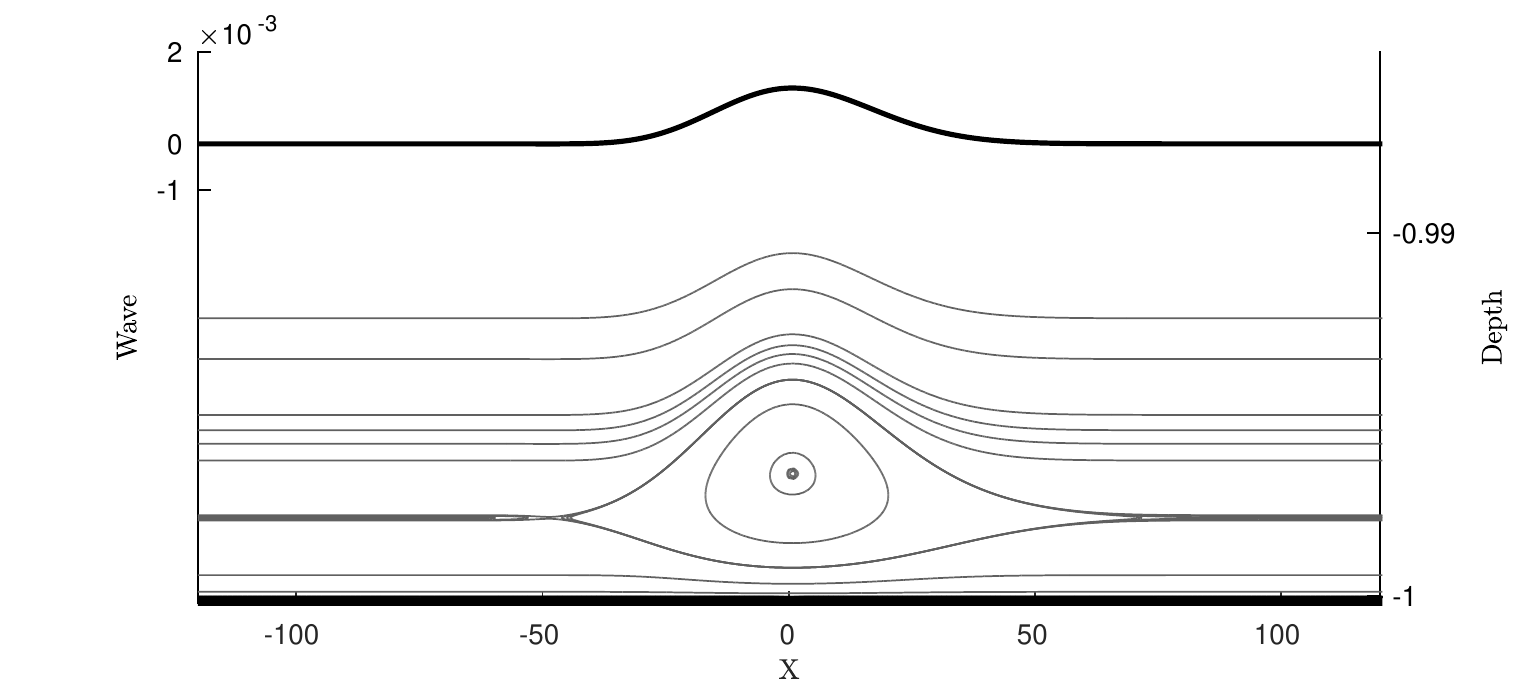}
	\caption{ Effective phase portrait using  $\psi_T(30)$. The left $y$-axis has values on the scale of the free surface, depicted at the top in a thicker dark line.
		The right $y$-axis has values on the  depth scale,   so that we can see the closed streamlines near the bottom.
	}
	\label{RetratoPressao}
\end{figure}

Based on the discussion above,   in figure   \ref{RetratoPressao}  we plot the streamlines for $\psi_T(30)$. We have a single cat eye (recirculation) region,
traveling with the wave speed. A critical layer divides a region into a rightgoing  and a leftgoing flow. 
As in earlier works, the critical layer  goes through the recirculation region. But in this case the critical layer then becomes a sharp interface emerging from 
what would have been a saddle point in the usual cat eye formation. The velocity field is continuous so there is no velocity jump crossing
the stagnation segment. Above the stagnation segment the flow is to the left while immediately below the interface the flow is to the right. 
In the present example we have an isolated  
stagnation point in the form of a center, in the middle of the recirculation
region, and stagnation segments connected to the isolated Kelvin cat eye.
This is a scenario not reported before, in particular for a non-stationary wave regime.

\begin{figure}
	\centering
	\includegraphics[width=0.9\textwidth]{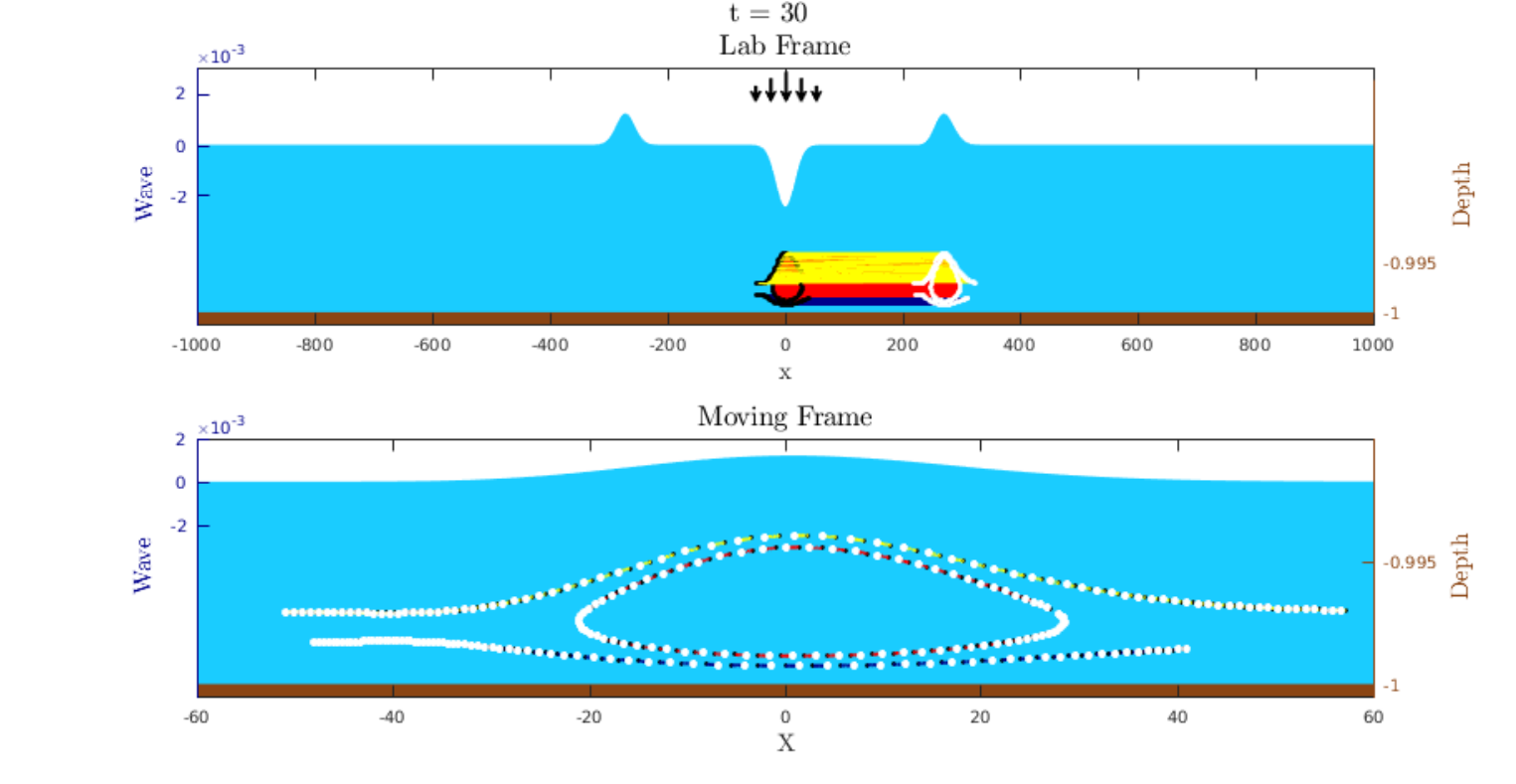}
	\noindent\rule{0.9\textwidth}{0.4pt}
	\includegraphics[width=0.9\textwidth]{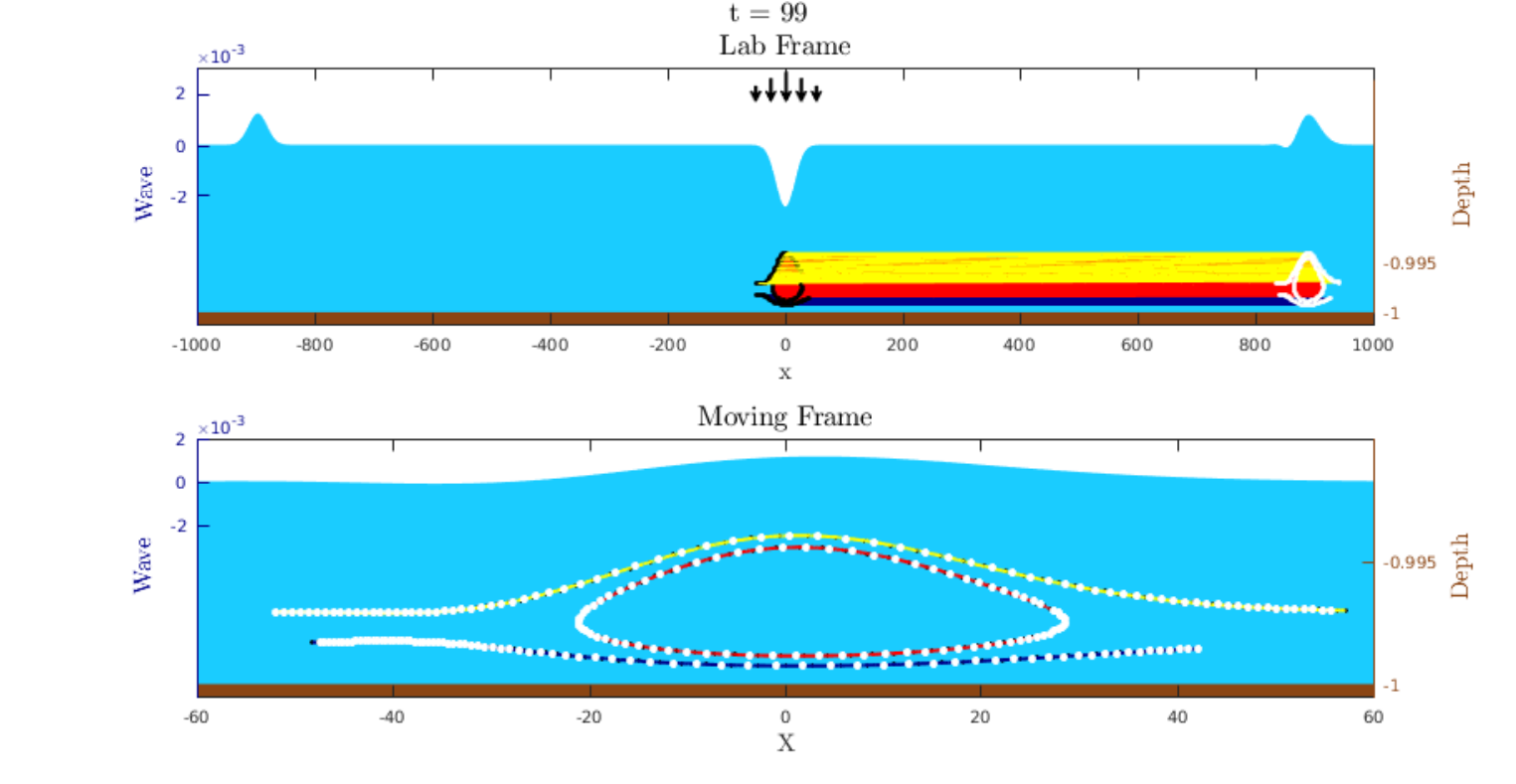}
	\caption{Continuation of figure \ref{OndaPressao}. The top two pictures show the tracers following the expected structure. The bottom  picture indicates that
		after a large propagation distance our cloud of markers has fully traced the Kelvin cat eye. 
		An animation (video03) can be found in the supplementary
material of this article.
			}
	\label{OndaPressao2}
\end{figure}

%

\section{Conclusions}\label{conclusao}
We have studied non-stationary rotational surface waves, in the presence of constant vorticity. We consider the regime where the Kelvin cat eye 
structure can be observed. In the linear regime one has a very narrow structure near the bottom. The accurate numerical method captured all details of the 
cat eye structure and the respective critical layer separating left from rightgoing streams.
Particle-trajectories were computed and visualized by  evolving the respective submarine dynamical system with a cloud of tracers. 
Waves were generated from either the current-topography interaction or by a surface pressure distribution suddenly imposed.

\section{Acknowledgements}
The work of M.F. was supported in part by CNPq-Science without Borders under (PDE) 200920/2014-6 and by CNPq-Cotas do Programa 
de P{\' o}s-Gradua{\c c}{\~ a}o (GM/GD)	140773/2014-2.
The work of A.N. was supported in part by CNPq under (PQ-1B) 304671/2017-7 and FAPERJ Cientistas do Nosso Estado project no. E-26/203.023/2017. 
R.R.-Jr  is grateful to IMPA for  the research support provided during the Summer Program of 2018  and to University of Bath for extended visit to the Department of Mathematical Sciences. 
A.N. and R.R.-Jr  thank Prof. Robin Johnson for his comments during theirs  
visits to Vienna. A.N. is grateful to the Erwin Schr\" odinger 
International Institute for Mathematics and Physics, for the support and hospitality 
during the ESI Programme on ``Mathematical Aspects of Physical Oceanography".
R.R.-Jr thanks Prof. Adrian Constantin for his support and hospitality during his
visit to the Department of Mathematics of the University of Vienna.

\bibliographystyle{abbrv}

\end{document}